\documentclass[pdflatex,sn-mathphys]{sn-jnl}% Math and Physical Sciences Reference Style
\jyear{2021}%

\theoremstyle{thmstyleone}%
\theoremstyle{thmstyletwo}%

\theoremstyle{thmstylethree}%

\usepackage{derivative}

\usepackage{xr}
\makeatletter

\newcommand*{\addFileDependency}[1]{% argument=file name and extension
\typeout{(#1)}% latexmk will find this if $recorder=0
\@addtofilelist{#1}
\IfFileExists{#1}{}{\typeout{No file #1.}}
}\makeatother

\newcommand*{\myexternaldocument}[1]{%
\externaldocument{#1}%
\addFileDependency{#1.tex}%
\addFileDependency{#1.aux}%
}
\myexternaldocument{SI}

\usepackage{subcaption}
\usepackage[section]{placeins}
\usepackage{tabularx}
\usepackage{multirow}
\usepackage{CJKutf8}

\usepackage{float}
\usepackage{etoolbox}
\makeatletter
\patchcmd{\ps@headings}
{\hbox to \hsize{\hfill Springer Nature 2021 \LaTeX\ template\hfill}}
{\hbox to \hsize{}}
{}
{}
\patchcmd{\ps@titlepage}
{\hbox to \hsize{\hfill Springer Nature 2021 \LaTeX\ template\hfill}}
{\hbox to \hsize{}}
{}
{}
\makeatother
\makeatletter
\patchcmd{\ps@headings}
{\hbox to \hsize{\hfill Springer Nature 2021 \LaTeX\ template\hfill}}
{\hbox to \hsize{}}
{}
{}
\patchcmd{\ps@headings}
{\hbox to \hsize{\hfill Springer Nature 2021 \LaTeX\ template\hfill}}
{\hbox to \hsize{}}
{}
{}
\patchcmd{\ps@titlepage}
{\hbox to \hsize{\hfill Springer Nature 2021 \LaTeX\ template\hfill}}
{\hbox to \hsize{}}
{}
{}
\makeatother

\begin{document}

\begin{CJK*}{UTF8}{gbsn}

\title[FuXi-S2S]{A machine learning model that outperforms conventional global subseasonal forecast models}


\author[1,2]{\fnm{Lei} \sur{Chen}}\email{cltpys@163.com}
\equalcont{These authors contributed equally to this work.}
\author[1]{\fnm{Xiaohui} \sur{Zhong}}\email{x7zhong@gmail.com}
\equalcont{These authors contributed equally to this work.}

\author*[1]{\fnm{Hao} \sur{Li}}\email{lihao$\_$lh@fudan.edu.cn}
\equalcont{These authors contributed equally to this work.}

\author[3]{\fnm{Jie} \sur{Wu}}\email{wujie@cma.gov.com}
\equalcont{These authors contributed equally to this work.}

\author*[3,4]{\fnm{Bo} \sur{Lu}}\email{bolu@cma.gov.cn}

\author[5]{\fnm{Deliang} \sur{Chen}}\email{deliang@gvc.gu.se}

\author[6]{\fnm{Shang-Ping} \sur{Xie}}\email{sxie@ucsd.edu}

\author[7,8,9]{\fnm{Libo} \sur{Wu}}\email{wulibo@fudan.edu.cn}

\author[3]{\fnm{Qingchen} \sur{Chao}}\email{chaoqc@cma.gov.cn}

\author[1]{\fnm{Chensen} \sur{Lin}}\email{linchensen@fudan.edu.cn}

\author[1]{\fnm{Zixin} \sur{Hu}}\email{huzixin@fudan.edu.cn}

\author*[2,1]{\fnm{Yuan} \sur{Qi}}\email{qiyuan@fudan.edu.cn}

\affil[1]{\orgdiv{Artificial Intelligence Innovation and Incubation Institute}, \orgname{Fudan University}, \orgaddress{\city{Shanghai}, \postcode{200433}, \country{China}}}

\affil[2]{\orgname{Shanghai Academy of Artificial Intelligence for Science}, \orgaddress{\city{Shanghai}, \postcode{200232}, \country{China}}}

\affil[3]{\orgdiv{China Meteorological Administration Key Laboratory for Climate Prediction Studies}, \orgname{National Climate Center}, \orgaddress{\city{Beijing}, \postcode{100081}, \country{China}}}

\affil[4]{\orgname{Xiong'an Institute of Meteorological Artificial Intelligence}, \orgaddress{\city{Xiong'an}, \country{China}}}

\affil[5]{\orgname{University of Gothenburg}, \orgaddress{\country{Sweden}}}

\affil[6]{\orgdiv{Scripps Institution of Oceanography}, \orgname{University of California San Diego}, \orgaddress{\country{USA}}}

\affil[7]{\orgdiv{School of Data Science}, \orgname{Fudan University}, \orgaddress{\city{Shanghai}, \postcode{200433}, \country{China}}}

\affil[8]{\orgdiv{Institute for Big Data}, \orgname{Fudan University}, \orgaddress{\city{Shanghai}, \postcode{200433}, \country{China}}}

\affil[9]{\orgdiv{MOE Laboratory for National Development and Intelligent Governance}, \orgname{Fudan University}, \orgaddress{\city{Shanghai}, \postcode{200433}, \country{China}}}

\abstract{Skillful subseasonal forecasts are crucial for various sectors of society but pose a grand scientific challenge. Recently, machine learning based weather forecasting models outperform the most successful numerical weather predictions generated by the European Centre for Medium-Range Weather Forecasts (ECMWF), but have not yet surpassed conventional models at subseasonal timescales. This paper introduces FuXi Subseasonal-to-Seasonal (FuXi-S2S), a machine learning model that provides global daily mean forecasts up to 42 days, encompassing five upper-air atmospheric variables at 13 pressure levels and 11 surface variables. FuXi-S2S, trained on 72 years of daily statistics from ECMWF ERA5 reanalysis data, outperforms the ECMWF's state-of-the-art Subseasonal-to-Seasonal model in ensemble mean and ensemble forecasts for total precipitation and outgoing longwave radiation, notably enhancing global precipitation forecast. The improved performance of FuXi-S2S can be primarily attributed to its superior capability to capture forecast uncertainty and accurately predict the Madden–Julian Oscillation (MJO), extending the skillful MJO prediction from 30 days to 36 days. Moreover, FuXi-S2S not only captures realistic teleconnections associated with the MJO, but also emerges as a valuable tool for discovering precursor signals, offering researchers insights and potentially establishing a new paradigm in Earth system science research.}

%A unique feature of FuXi-S2S is its perturbation module, which introduces flow-dependent perturbations in hidden features, enhancing forecast uncertainty estimation and subseasonal forecast skill. 


\maketitle

\section{Introduction}

Subseasonal forecasting, which predicts weather patterns from 2 to 6 weeks in advance, bridges a critical gap between short-term weather forecasts, typically up to 15 days, and longer-term climate forecasts that extend to seasonal and longer timescales \cite{board2016next}. Forecasting at this intermediate subseasonal timescale is indispensable for a variety of applications, including agricultural planning, disaster preparedness, mitigating impacts of extreme events such as heatwaves, droughts, floods, and cold spells, and water resource management \cite{White2017,pegion2019subseasonal,White2022,domeisen2022}. Despite its significant socioeconomic benefits, subseasonal forecasting has historically not received sufficient attention compared to medium-range weather and climate predictions. This gap existed because accurate subseasonal forecasts were once considered nearly impossible. Subseasonal forecasts are particularly challenging as they rely on both atmospheric initial conditions, essential in short-term weather forecasts, and boundary conditions at the Earth's surface, key to seasonal and climate forecasts \cite{Lorenz1979,mariotti2018progress}. However, neither of these condition provides sufficient predictability, leaving subseasonal forecasts in a so-called predictability desert. Despite these challenges, recent advances in both physical and statistical modeling have enabled the regular production of subseasonal forecasts globally. Nonetheless, there remains a ongoing, strong demand for their further development to support informed decision-making across various sectors.

%每段话第一句知道这段话希望讲什么：现在的预报技巧还不好，讲出现在的方法（NWP models 计算成本太高不允许有很多集合预报、集合扰动的生成方式很多不是flow-depedent, postprocessing, direct forecasting with ML models流依赖扰动没有）和存在的问题，包括传统方法的问题、订正的问题、直接预报的问题，合并下面三段话
Developing an ensemble prediction system (EPS) based on traditional physics-based numerical weather prediction (NWP) models is a widely acknowledged and effective method for enhancing subseasonal forecast accuracy \cite{Weyn2021sub,han2023ensemble}. Major forecasting centers have implemented such EPS for subseasonal forecasts \cite{Vitart2014Sub,saha2014ncep,vitart2017subseasonal,pegion2019subseasonal}. However, these systems often exhibit considerable biases \cite{nowak2017sub,Monhart2018,hwang2019improving,vitart2022,Mouatadid2023adaptive}, particularly in predicting extreme events \cite{domeisen2022advances}. The two primary challenges in this field are ensuring an adequate ensemble size within computational constraints and designing ensemble perturbations that accurately reflect uncertainty in key atmospheric and oceanic variability \cite{Demaeyer2022}. Enlarging the ensemble size is beneficial for forecast performance \cite{buizza1999stochastic,Buizza2019,leutbecher2019ensemble}, but the substantial computational costs typically limit ensemble sizes to between 4 and 51 members across 11 international forecasting centers \cite{vitart2017subseasonal}. Given these computational limitations, machine learning model emerges as a promising alternative for direct subseasonal forecasting \cite{cohen2019s2s}. Machine learning models have the advantages of significantly higher computational efficiency, facilitating the generation of a large number of ensemble members which are crucial for prediction skill and reliability \cite{richardson2001measures}. Recent advancements in machine learning for medium-range weather forecasting \cite{hu2022swinvrnn,pathak2022fourcastnet,lam2022graphcast,bi2022panguweather,chen2023fuxi,fuxi_extreme2023,nguyen2023scaling} have demonstrated that machine learning models can outperform the high-resolution forecasts (HRES) generated by the European Centre for Medium-Range Weather Forecasts (ECMWF), widely considered as the most accurate global weather forecasts \citep{ECMWF2021}.

%We can not solve our problems with the same level of thinking that created them. 

%limited variables, and no effective ensemble perturbations
%为什么别人没报好，别人没考虑什么（全面的多变量，集合方式不太好），而我们预报的好 flow-dependent，可以称为我们的是state-of-the-art model
Machine learning models have achieved made significant strides in medium-range weather forecasting and seasonal forecasting \cite{wang2024coupled}, but their success in subseasonal forecasting has been less pronounced \cite{Weyn2021sub,He2021,kiefer2023can}. This shortfall  primarily stems from the limited range of variables incorporated into the models, and more importantly, from the inadequate methods employed for ensemble generation. Conventional machine learning techniques for ensemble forecasting, such as introducing random perturbations into initial conditions and altering model structures, overlook the background flow and consequently leads to rapid reduction in ensemble spread. The inadequate representation of the complexities limits the performance of these prior machine learning based subseasonal forecasting models, which does not yet rival that of traditional EPS based on NWP models. To overcome these challenges, we introduce the FuXi Subseasonal-to-Seasonal (FuXi-S2S) model, representing a significant advancement in machine learning for subseasonal forecasting. This model is designed to generate global daily mean forecasts for 42 days from initialization. Unlike previous models that incorporated a limited set of variables, it incorporates a comprehensive suite of variables, instead of a couple of variables in previous models: 5 upper-air atmospheric variables at 13 pressure levels and 11 surface variables. Furthermore, it features a innovative perturbation module specifically designed to generate flow-dependent perturbations for ensemble forecasting. This module leverages vast amounts of historical data to learn probability distributions, thereby introducing flow-dependent perturbations directly into the model's hidden features. Compared to conventional NWP ensemble forecasting methods, which often struggle with constructing initial condition perturbations due to the complexities of multivariate interactions and the need to maintain dynamic balance and ensemble spread in simulations \cite{Molteni1996}, our approach of introducing perturbations directly into the model's latent space, presenting a novel and effective alternative. This perturbation module significantly enhances the performance of the FuXi-S2S forecasts, as demonstrated in Supplementary Figure \ref{flow_dependent}. More details about the FuXi-S2S model architecture are available in Section 4.

Remarkably, FuXi-S2S outperforms the ECMWF Subseasonal to Seasonal (S2S) ensemble, which is recognized as the most skillful S2S modeling system, in producing both the ensemble mean and probabilistic forecasts \cite{de2019global,domeisen2022}. Its efficacy is particularly evident in extreme total precipitation ($\textrm{TP}$) forecasting, as exemplified by its accurate forecasts for the 2022 Pakistan floods. Such capability is closely related to FuXi-S2S's improved prediction of the Madden–Julian Oscillation (MJO) \cite{madden1971detection,madden1972description}, a key driver of global climate patterns, extending the skillful MJO prediction from 30 days to 36 days. These results further confirm that the notable improvement in FuXi-S2S's performance can be primarily attributed to the innovative perturbation module for ensemble generation. Another promising result is the ability of the FuXi-S2S model to identify potential precursor signals to physical processes. Beyond mere accuracy, in many applications involving machine learning forecasts, it is imperative to understand and validate the decision-making mechanisms of these models. Such understanding not only leads to enhanced trust in the models' predictions but also increases the likelihood of implementing effective actions, particularly in mitigating the risks associated with extreme events. Therefore, interpreting machine learning models to align their reasoning with established knowledge becomes crucial. Recent developments in explainable machine learning (XML) \cite{LPR2015,McGovern2019,molnar2020,mamalakis2020,toms2021testing,rasp2021data} methods have facilitated this interpretation. This study delves into the 2022 Pakistan floods, investigating the FuXi-S2S model's predictions to identify key geographic regions that significantly impact its predictive accuracy. This is achieved through the generation and analysis of saliency maps \cite{simonyan2013deep}, wherein the identified regions in close alignment with insights from previous studies \cite{dunstone2023windows}. Therefore, we argue that FuXi-S2S transcends traditional NWP models in terms of accuracy and speed, potentially unveiling previously unrecognized processes within Earth's system in subseasonal forecasting \cite{faghmous2014big,karpatne2017theory}.

\section{Results}

This study conducts a thorough evaluation of the 51-member FuXi-S2S forecasts by analyzing testing data spanning from 2017 to 2021. It compares the performance of FuXi-S2S with that of the 11-member ECMWF S2S reforecasts from the model cycle C47r3 over the same period. The analysis primarily focuses on average forecasts for week 3 (days 15-21), week 4 (days 22-28), week 5 (days 29-35), and week 6 (days 36-42), weeks 3-4, and weeks 5-6. The evaluation employs a comprehensive set of metrics, including deterministic metrics for the ensemble mean, probabilistic metrics for all ensemble members, prediction skills specific for MJO forecasts, and tailored assessments for extreme events, notably the 2022 Pakistan floods. Furthermore, the study explores the underlying processes driving the FuXi-S2S model's predictions for the 2022 Pakistan floods. This is accomplished by generating and analyzing the saliency maps, which provides profound insights into the model's predictive processes.

Additional evaluations, including an analysis of energy spectra \cite{chattopadhyay2023longterm}, are available in the supplementary material.
%Additionally, the performance of FuXi-S2S forecasts for the 2022 Pakistan floods is evaluated and compared to the 51-member ECMWF S2S real-time forecasts from model cycle C47r3. 

\subsection{Deterministic metrics}

\begin{figure}[h]
    \centering
    \includegraphics[width=\linewidth]{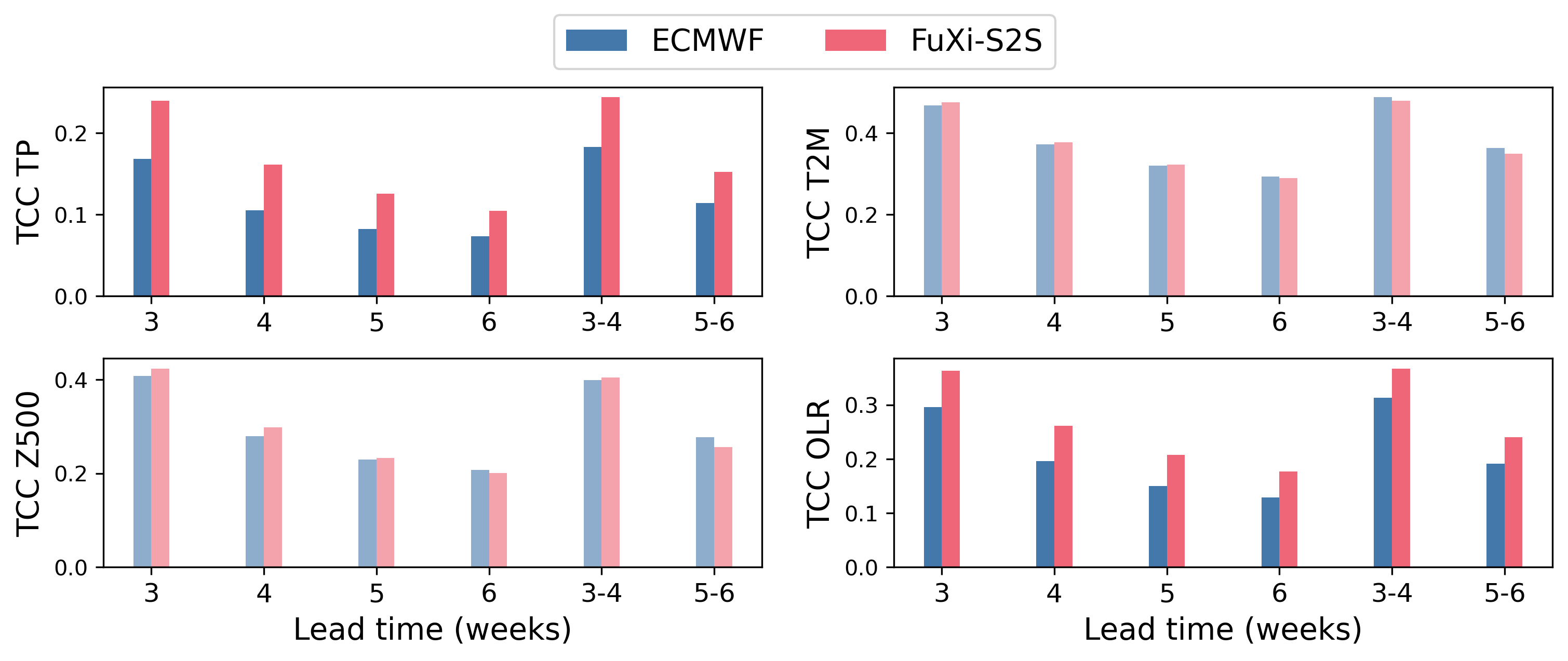}
    \caption{Comparison of globally-averaged and latitude-weighted temporal anomaly correlation coefficient ($\textrm{TCC}$) of the ensemble mean between ECMWF subseasonal-to-seasonal (S2S) reforecasts (in blue) and FuXi-S2S forecasts (in red) for total precipitation ($\textrm{TP}$), 2-meter temperature ($\textrm{T2M}$), geopotential at 500 hPa ($\textrm{Z500}$), and outgoing longwave radiation ($\textrm{OLR}$). Rows 1 and 2 represent the performance across these variables, utilizing all testing data from the period spanning from 2017 to 2021. A bootstrapping approach, repeated 1000 times, is used for significance testing. When the FuXi-S2S forecasts fail to show a statistically significant improvement over the ECMWF S2S reforecasts at the 97.5\% confidence level, a pale color scheme is used to denote these results.}
    \label{TCC}    
\end{figure}

This subsection compares the performance of ensemble mean forecasts from FuXi-S2S and ECMWF S2S based on deterministic metrics. Figure \ref{TCC} presents the globally-averaged and latitude-weighted temporal anomaly correlation coefficient ($\textrm{TCC}$) for both FuXi-S2S and ECMWF S2S, considering four variables: ${\textrm{TP}}$, 2-meter temperature (${\textrm{T2M}}$), geopotential at 500 hPa (${\textrm{Z500}}$), and outgoing longwave radiation (${\textrm{OLR}}$), across forecast lead times of 3, 4, 5, 6, 3-4, and 5-6 weeks. Significance testing is conducted as described in Section 4.4. When the FuXi-S2S forecasts do not show a statistically significant improvement over the ECMWF S2S reforecasts, these are indicated with a pale color scheme. It is evident that the ensemble mean forecasts from FuXi-S2S significantly outperform ECMWF S2S for ${\textrm{TP}}$ and ${\textrm{OLR}}$, but not for ${\textrm{T2M}}$ and ${\textrm{Z500}}$. The analysis is based on the averaged $\textrm{TCC}$ computed from all testing data spanning the period from 2017 to 2021. The FuXi-S2S forecasts generally demonstrate higher TCC values than the ECMWF S2S reforecasts for ${\textrm{TP}}$ and ${\textrm{OLR}}$ at all lead times, while comparable $\textrm{TCC}$ values for ${\textrm{Z500}}$ and ${\textrm{T2M}}$. Specifically, regarding ${\textrm{Z500}}$, the FuXi-S2S forecasts are superior to the ECMWF S2S reforecasts at lead times of 3, 4, 5, and 3-4 weeks, and have inferior performance at lead times of 6 and 5-6 weeks. 

Supplementary Figure \ref{TCC_spatial} provides the spatial distributions of temporally-averaged $\textrm{TCC}$ for both ECMWF S2S and FuXi-S2S, along with the differences in $\textrm{TCC}$ between FuXi-S2S and ECMWF S2S for $\textrm{TP}$, $\textrm{T2M}$, $\textrm{Z500}$, and $\textrm{OLR}$ forecasts at lead times of 3-4 and 5-6 weeks, respectively. The spatial distributions of $\textrm{TCC}$ reveal considerably higher values over tropics, and greater values over oceans than over land. The $\textrm{TCC}$ differences are described in red (positive values), blue (negative values), and white (zero values) patterns, suggesting whether FuXi-S2S's performance is superior, inferior, or equivalent to ECMWF S2S, respectively. Overall, FuXi-S2S demonstrates positive $\textrm{TCC}$ differences for ${\textrm{TP}}$ and ${\textrm{OLR}}$ in most regions worldwide, consistent with the findings presented in Figure \ref{TCC}. Moreover, FuXi-S2S also outperforms ECMWF in a majority of extra-tropical regions for both $\textrm{T2M}$ and $\textrm{Z500}$, although its performance is generally less skilful in the tropical areas.

\subsection{Probabilistic metrics}

Deterministic metrics, evaluated using the ensemble mean, exhibit limited predictive skill, with $\textrm{TCC}$ values below 0.5 for all subseasonal forecast lead times. Therefore, ensemble forecasts are essential for detecting predictable signals at subseasonal timescales. 
%This subsection focuses on examining ensemble forecast metrics, comparing the FuXi-S2S and ECMWF S2S models. 

\begin{figure}[h]
    \centering
    \includegraphics[width=\linewidth]{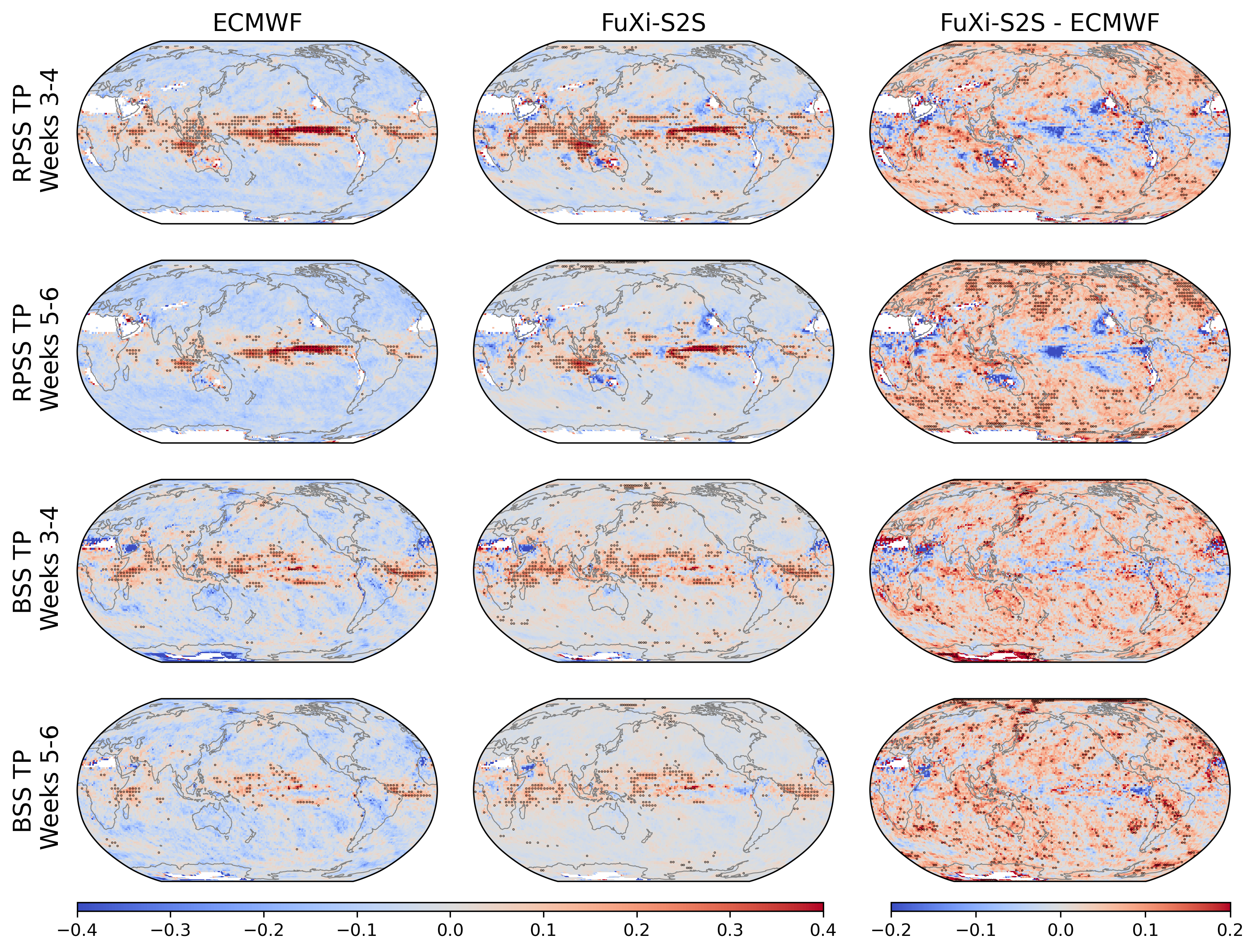}
    \caption{Maps displaying the average Ranked Probability skill Score ($\textrm{RPSS}$) (first and second rows) and Brier Skill Score ($\textrm{BSS}$) (third and fourth rows) without latitude weighting, comparing ECMWF subseasonal-to-seasonal (S2S) (first column) and FuXi-S2S (second column) forecasts. Additionally, the third column depicts the difference in $\textrm{RPSS}$ and $\textrm{BSS}$ between FuXi-S2S and ECMWF S2S for total precipitation ($\textrm{TP}$) at forecast lead times of weeks 3-4 (first and third rows) and weeks 5-6 (second and fourth rows), utilizing all testing data from 2017 to 2021. Red contour lines in the first and second columns indicate areas with positive values of $\textrm{RPSS}$ and $\textrm{BSS}$. Stippling on the map denotes areas where the skill score is statistically significant at the 97.5\% confidence level. Specifically, in columns 1 and 2, stippling indicates regions where the skill scores of the ECMWF S2S and FuXi-S2S models significantly surpasses those of climatology. In column 3, stippling highlights areas where the FuXi-S2S model significantly outperforms the ECMWF S2S.}
    \label{RPSS_BSS_spatial}    
\end{figure}

The first two rows of Figure \ref{RPSS_BSS_spatial} present the spatial distribution of the temporally-averaged ranked probability skill score ($\textrm{RPSS}$) \cite{epstein1969scoring,Wilks2011} for ECMWF S2S and FuXi-S2S, as well as the $\textrm{RPSS}$ differences between FuXi-S2S and ECMWF S2S for $\textrm{TP}$ forecasts over 3-4 and 5-6 week lead times. This analysis utilizes $\textrm{RPSS}$ data which are temporally averaged from 2017 to 2021. The red contour lines in the first and second columns highlight areas with positive $\textrm{RPSS}$ values, which indicate more skillful prediction than climatology forecast can be obtained over these areas. Notably, FuXi-S2S predicts more areas with positive $\textrm{RPSS}$ values than ECMWF S2S. The color coding in the right panels of Figure \ref{RPSS_BSS_spatial} (red, blue, and white) indicates regions where FuXi-S2S performs better, worse, or equivalently compared to ECMWF S2S, respectively. The global distribution of $\textrm{RPSS}$ suggests that both ECMWF S2S and FuXi-S2S primarily exhibit skill in tropical regions, whereas they lack skill in the extra-tropics compared to climatology. In contrast, $\textrm{RPSS}$ demonstrates positive values (depicted in red color) in tropical regions, indicating enhanced predictive skills relative to climatology. Moreover, the $\textrm{RPSS}$ values are notably higher over oceans compared to land areas. Predominantly, FuXi-S2S demonstrates nearly global positive $\textrm{RPSS}$ differences for ${\textrm{TP}}$, except in some tropical regions where both models have quite high $\textrm{RPSS}$ values. Compared to ECMWF S2S, whose skillful predictions are primarily confined to tropical ocean areas, FuXi-S2S demonstrates the capability of skillful predictions over more extra-tropical regions, such as East Asia, the North Pacific and the Arctic.

The latitude-weighted $\textrm{RPSS}$ for the same 4 variables as in Figure \ref{TCC} over forecast lead times of 3, 4, 5, 6, 3-4, and 5-6 weeks are given in Supplementary Figure \ref{RPSS_area_bar}. FuXi-S2S shows higher \textrm{RPSS} values than ECMWF S2S across most regions for all the examined variables: ${\textrm{TP}}$, ${\textrm{T2M}}$, ${\textrm{Z500}}$, and ${\textrm{OLR}}$. This superiority is especially noticeable in extratropical averages. However, in the tropics, ECMWF S2S outperforms FuXi-S2S at lead times of 3 to 6 weeks for one-week averages, whereas FuXi-S2S surpasses ECMWF S2S for two-week averages. This discrepancy in performance likely arises from the fact that one-week averages filter out variability with periods shorter than two weeks, while two-week averages attenuate variability with periods shorter than four weeks. Thus, the skill differences between the one-week and two-week averages may reflect FuXi-S2S's enhanced ability in capturing lower-frequency variability. Furthermore, a previous study \cite{de2019global} suggests that dynamical S2S models, particularly ECMWF S2S, demonstrate improved performance in the central-eastern Pacific, potentially due to their effective simulation of the realistic air-sea interactions in these regions.

%In contrast, for ${\textrm{T2M}}$, FuXi-S2S tends to perform worse compared to ECMWF S2S, except in some oceanic areas where ECMWF S2S is less effective, as indicated by the white or blue areas in its $\textrm{RPSS}$. 

\subsection{Extreme forecast}

A primary target of subseasonal forecasts is extreme weather events, to better prepare for disasters like droughts and floods. This subsection focuses on the prediction skills for extreme precipitation events. Such events are identified when $\textrm{TP}$ exceeds the 90th climatological percentile, a threshold that varies based on grid location, forecast initialization time, and forecast lead time.

The last two rows of Figure \ref{RPSS_BSS_spatial} show the spatial distributions of the temporally-averaged Brier Skill Score ($\textrm{BSS}$) \cite{Wilks2011} for the extreme precipitation events, for ECMWF S2S and FuXi-S2S, and their differences over 3-4 and 5-6 week lead times. Similar to spatial pattern of $\textrm{RPSS}$, FuXi-S2S generally exhibts more regions with positive values of $\textrm{BSS}$ than ECMWF S2S, suggesting more areas with skill relative to climatological forecasts. Similar to spatial pattern of $\textrm{RPSS}$, the $\textrm{BSS}$ values are considerably higher over oceans than over land and decrease from lower latitudes to higher latitudes. Predominantly, the $\textrm{BSS}$ differences favor FuXi-S2S in $\textrm{TP}$ over land and in extra-tropical regions, marked by widespread red patterns. This suggests FuXi-S2S's dominance over ECMWF S2S in predicting extreme $\textrm{TP}$ across land and extra-tropics, which is of great importance for disaster preparedness and early warning. 
%Furthermore, a significant portion of tropical ocean exhibits negative $\textrm{T2M}$ differences, consistent with the statistical findings for both extra-tropics and tropics.

Supplementary Figure \ref{BSS_area_bar} compares the latitude-weighted $\textrm{BSS}$ between FuXi-S2S and ECMWF S2S, focusing on $\textrm{TP}$, $\textrm{T2M}$, $\textrm{Z500}$, and $\textrm{OLR}$ in five geographical regions: global, in the extra-tropics (90\textdegree S - 30\textdegree S and 30\textdegree N - 90\textdegree N), in the tropics (30°S - 30°N), over land, and over the ocean. Globally, FuXi-S2S outperforms ECMWF S2S in terms of $\textrm{BSS}$ for $\textrm{TP}$, $\textrm{T2M}$, and $\textrm{OLR}$. Notably, in contrast to ECMWF S2S, which exhibits consistently negative globally-averaged \textrm{BSS} values for $\textrm{TP}$ across all lead times, FuXi-S2S demonstrates positive values for forecast lead times of 3, 3-4 and 5-6 week. In the extra-tropical regions, though the \textrm{BSS} scores are relatively lower in comparison to the global average, FuXi-S2S consistently exhibits superior performance compared to ECMWF S2S across all four variables. A similar pattern emerges in tropical regions, where FuXi-S2S demonstrates superior performance over ECMWF S2S for $\textrm{TP}$ and $\textrm{OLR}$, while achieving comparable accuracy in $\textrm{T2M}$ and $\textrm{Z500}$. Over land areas, FuXi-S2S demonstrates consistently higher \textrm{BSS} values for $\textrm{TP}$ and $\textrm{T2M}$, suggesting its superior ability to provide more accurate forecasts of extreme rainfall and high temperatures compared to ECMWF S2S.

\subsection{MJO forecast}

Recent studies have demonstrated the importance of accurately modeling various sources of subseasonal predictability, particularly the MJO \cite{vitart2017subseasonal,vitart2018sub,merryfield2020}, for improving subseasonal prediction skills. The MJO has a significant impact on global weather and climate, serving as a primary source of predictability at subseasonal timescales due to its quasi-periodic nature \cite{Chidong2005,zhang2013madden,zhang2013cracking,neena2014predictability}. Accurate MJO prediction is essential for reliable subseasonal predictions. Although current state-of-the-art dynamical forecasts can predict the MJO up to 3-4 weeks in advance, this falls short of the theoretical potential predictability of approximately 6-7 weeks \cite{neena2014predictability,kim2018,jiang2020fifty}. In recent years, increasing efforts have focused on applying machine learning models to improve MJO forecasts, either by post-processing dynamical forecasts \cite{Jie2021,silini2022improving,Kim2023} or through direct forecasting \cite{silini2021machine,toms2021testing,Delaunay2022}. However, only improving MJO predictions with machine learning models does not inherently ensure improved forecasts of related weather phenomena, such as tropical cyclones and monsoons, which also depend on accurate predictions of various weather parameters by the model. Therefore, continuous improvement in forecasting models is essential for advancing subseasonal prediction capabilities. This section specifically examines the performance of our FuXi-S2S model in MJO forecasts, although it is not explicitly optimized for this purpose.

\begin{figure}[h]
    \centering
    \includegraphics[width=\linewidth]{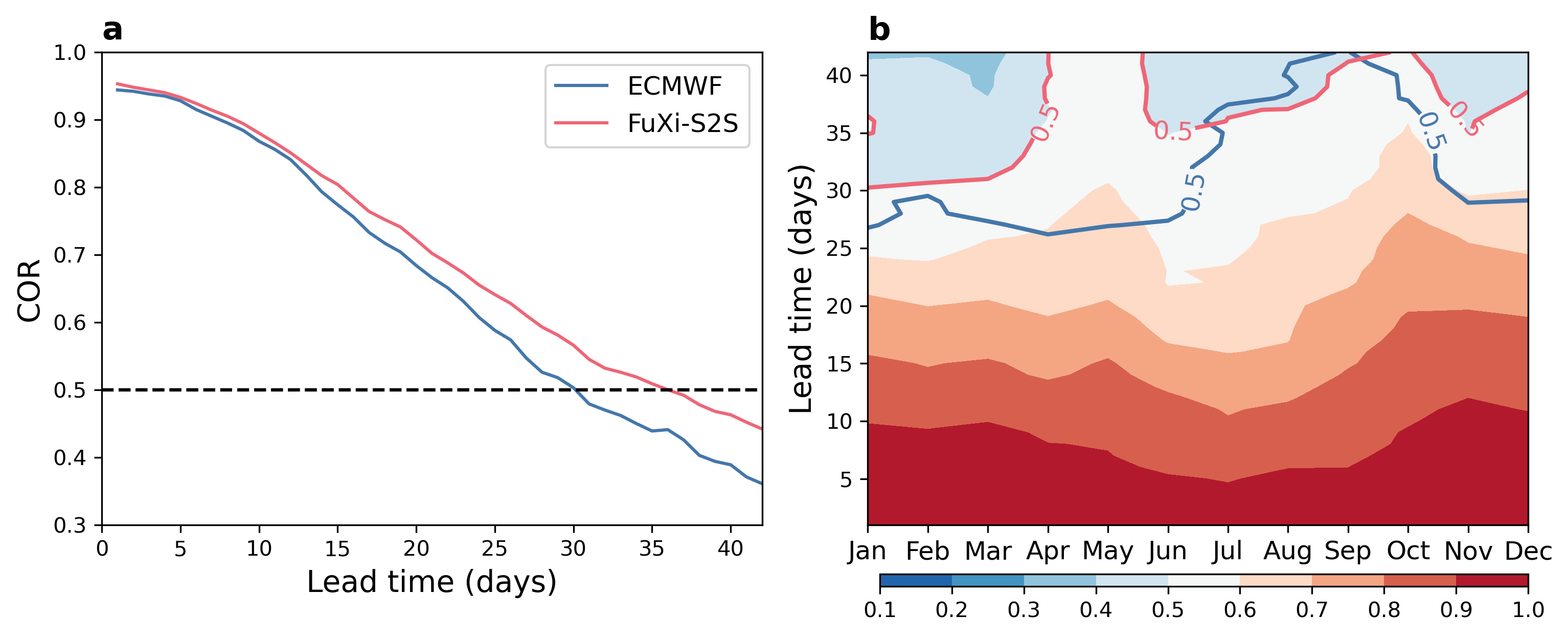}
    \caption{Comparison of real-time multivariate Madden–Julian Oscillation (MJO) (RMM) bivariate Correlation ($\textrm{COR}$) of the ensemble mean between ECMWF subseasonal-to-seasonal (S2S) reforecasts (in blue) and FuXi-S2S forecasts (in red) using all testing data from 2017 to 2021. $\textbf{a}$) Comparison of RMM bivariate $\textrm{COR}$ as a function of forecast lead times. Dashed black line signifies the prediction skill threshold of $\textrm{COR}$=0.5. $\textbf{b}$) The RMM bivariate $\textrm{COR}$ is depicted as a function of the month of initialization (x-axis) and forecast lead time (y-axis), with red and blue lines indicating the skillful MJO prediction days of ECMWF S2S (in blue) and FuXi-S2S (in red), respectively.}
    \label{MJO_skill}    
\end{figure}

In this study, we employed the real-time multivariate MJO (RMM) index \cite{wheeler2004all}, along with the commonly used metrics of bivariate correlation coefficient (COR), to evaluate the forecasting skill of the MJO. The RMM index used for verification was calculated using the Climate Prediction Center (CPC) OLR (CBO) data, in conjunction with the ERA5 zonal-wind component at 850 hPa and 200h Pa. Figure \ref{MJO_skill} presents the bivariate correlation ($\textrm{COR}$) skills of the RMM index for the ensemble mean of ECMWF S2S reforecasts and FuXi-S2S forecasts, averaged over the testing data spanning from 2017 to 2021. The results show a decrease in COR values as forecast lead times increase. Particularly, FuXi-S2S outperforms ECMWF S2S in MJO prediction, maintaining higher COR values for up to 42 days. When applying a $\textrm{COR}$ threshold of 0.5 to determine skillful MJO forecast, FuXi-S2S extends the skillful forecast lead time from 30 days to 36 days, surpassing the performance of ECMWF S2S. Furthermore, the MJO prediction skills also depend on the seasonal cycle, as illustrated in Figure \ref{MJO_skill}. Both FuXi-S2S and ECMWF S2S demonstrate higher MJO prediction skills in September and October. Additionally, FuXi-S2S exhibit superior skills compared to ECMWF S2S during the boreal spring and winter, with skillful predictions extending beyond 42 days in April and May, which is the longest forecast lead time achievable by the FuXi-S2S model. Moreover, Supplementary Figure \ref{mjo_amp_phase} presents the $\textrm{COR}$ and error for the amplitude and phase of the MJO. These are calculated using the ensemble mean of ECMWF S2S reforecasts and FuXi-S2S forecasts, averaged across over the 2017-2021 testing dataset. The results suggest that the FuXi-S2S model outperforms the ECMWF S2S model in predicting the MJO, primarily due to its superior capability in forecasting the MJO phase. Additionally, FuXi-S2S demonstrates smaller amplitude errors, suggesting it more accurately maintains the amplitude of MJO events.

%the skillful MJO prediction of FuXi-S2S exceeds 42 days, which is also the longest forecast lead time achievable by FuXi-S2S.

%In terms of RMSE, with a threshold value of 1.4 defining prediction skill, both models show longer periods of skillful prediction. 
%Notably, FuXi-S2S's RMSE remains below this threshold for the entire forecast time, while ECMWF S2S's RMSE falls below the RMSE threshold after 41 days.

A two-dimensional phase-space diagram is commonly used to characterize the phase and amplitude of the MJO, using the x-axis and y-axis to represent the first and second principal components of Empirical Orthogonal Functions (EOFs) (RMM1 and RMM2), respectively. Supplementary Figure \ref{RMM_composite} illustrates the forecast performance of four distinct MJO events with initialization dates of 27 June 2018, 3 November 2018, 18 April 2019, and 21 March 2021, as predicted by ECMWF S2S and FuXi-S2S. Data points on this two-dimensional phase-space diagram are plotted at 5-day intervals. The phase of the MJO is determined by the azimuth of the combined RMM indices 1 and 2 (RMM1 and RMM2), while its amplitude is represented by the radial distance from the origin. As visually shown in Supplementary Figure \ref{RMM_composite}, the counterclockwise movement of data points signifies the eastward propagation of MJO-associated convection, with the distance between successive points reflecting the propagation speed. In comparison to the observed MJO derived from CBO and ERA5 reanalysis data, both ECMWF S2S and FuXi-S2S exhibit slower propagation speeds and reduced amplitudes as the forecast lead time increases, particularly noticeable for MJO forecasts initialized on 21 March 2021. However, FuXi-S2S shows a more consistent alignment with observations across all MJO phases, especially in mitigating the negative amplitude biases in MJO forecasts when compared to ECMWF S2S.

The MJO originates from interactions of tropical convection and circulation but its effect is of global reach. Indeed, large TCC for $\textrm{Z500}$ over the extra-tropical Pacific is found along the path of the Pacific North/South American (PNA/PSA) \cite{wallace1981teleconnections,mo1987statistics} teleconnection pattern (Supplementary Figure \ref{TCC_spatial}, rows 6 and 7). Compared to ECMWF S2S, improved MJO forecast in FuXi-S2S elevates TCC for these teleconnection patterns, especially along the PSA wave train in the Southern Hemisphere. Furthermore, the MJO is critical for stimulating these important teleconnection patterns, significantly affecting extra-tropical anomalies. Therefore, the accurate representation of MJO-related teleconnections is imperative for effective subseasonal forecasts. Supplementary Figure \ref{z500} demonstrates that the FuXi-S2S model showcases enhanced skills in MJO prediction and realistic simulations of MJO teleconnections, which substantially contribute to its superior performance in subseasonal forecasts, particularly over extra-tropical regions.

This study highlights FuXi-S2S proficiency in predicting the MJO. We envision that FuXi-S2S could serve as a pivotal tool in investigating other primary modes of subseasonal variability, such as the Boreal Summer Intraseasonal Oscillation (BSISO) \cite{Zhu1993The3C}, North Atlantic Oscillation (NAO) \cite{walker1924correlations}, and East Asia-Pacific (EAP) pattern \cite{huang1987influence}. Additionally, it would be worthwhile to explore how the prescribed fixed sea surface temperature (${\textrm{SST}}$) or its absence impacts the forecast performance of the MJO. Savarin and Chen \cite{savarin2022pathways} demonstrated that either using a coupled atmosphere-ocean model or updating ${\textrm{SST}}$ with observed values is essential for accurately modeling the eastward propagation of the MJO. However, this analysis is beyond the scope of the current study and will be addressed in future research.

%Furthermore, Supplementary Figure 3 illustrates snapshots of daily $\textrm{OLR}$ anomalies, indicative of MJO-related convection, for ERA5 reanalysis data, the ensemble mean of ECMWF S2S reforecasts, and FuXi-S2S forecasts. Rows 1 to 8 represent eight forecast lead times, ranging from day 1 to day 36 in 5-day intervals (day 1, 6, 11, 16, 21, 26, 31, 36), all initialized from 21 April 2019. Specifically, day 1 corresponds to a MJO event transitioning from the end of phase 2 to the start of phase 3 in its life cycle. Both ECMWF S2S and FuXi-S2S forecast the geographical distribution of MJO OLR anomalies reasonably well. These models effectively capture the eastward movement of negative OLR anomalies from day 1 to day 26 (rows 1 to 6), covering phase 2 to phase 8. However, the models increasingly underestimate MJO OLR anomalies with longer lead times, a trend more and more evident from day 16 onward (rows 4 to 8). This observation aligns with the phase-space diagram in Figure 2. From day 31 (7th row), as a new MJO event develops over the Indian Ocean, both models demonstrate proficiency in predicting successive MJO occurrences. Nonetheless, despite FuXi-S2S more accurately predicting spatial distribution, both models considerably underestimate the intensity of the new MJO event compared to ERA5 reanalysis data.

\subsection{Prediction of the 2022 Pakistan floods}

In 2022, Pakistan experienced a series of exceptionally intense monsoon rainfall surges from early July to late August, resulting in total rainfall that reached a level approximately four standard deviations above the climatological mean \cite{hong2023causes}. This extreme rainfall event led to a significant humanitarian disaster, leaving over 2.1 million people homeless and resulting in 1,730 fatalities. According to the World Bank, the economic damages and losses exceeded USD 30 billion \cite{dunstone2023windows}. Consequently, it is important to assess the ability of subseasonal forecasts to predict such extreme rainfall events. 

\begin{figure}[h]
    \centering
    \includegraphics[width=\linewidth]{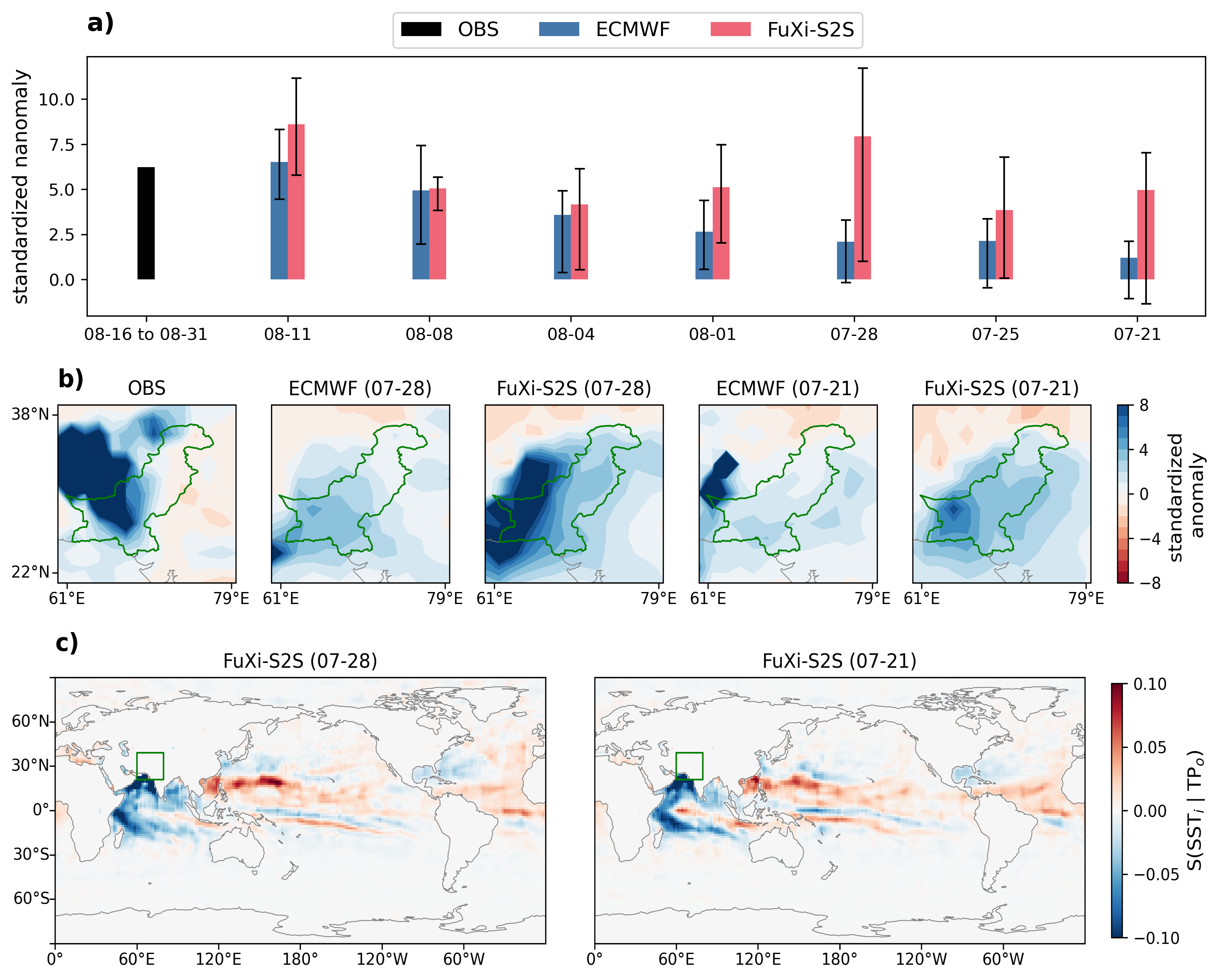}
    \caption{Comparative analysis for the 2022 Pakistan floods predictions between the ECMWF subseasonal-to-seasonal (S2S) and FuXi-S2S models as well as the precursor signals that contributed to accurate predictions by the FuXi-S2S model. Comparison of spatially and temporally averaged standardized total precipitation ($\textrm{TP}$) anomaly (a) over the two weeks from August 16th to August 31st, 2022, showcasing GPCP observations (in black) alongside predictions from ECMWF S2S real-time forecasts (in blue) and FuXi-S2S forecasts (in red), with initialization dates: August 11th (08-11, MM-DD), August 8th (08-08), August 4th (08-04), August 1st (08-01), July 28th (07-28), July 25th (07-25), and July 21st (07-21). The black lines on the bar of ECMWF S2S and FuXi-S2S forecasts represent the 25th and 75th percentiles. For the comparison of temporally averaged standardized $\textrm{TP}$ anomaly maps (b), the first column represents GPCP observations, while the second and third columns display predictions from ECMWF S2S and FuXi-S2S, respectively, both initialized on July 28th, and the fourth and fifth columns correspond to predictions from ECMWF S2S and FuXi-S2S, respectively, with an initialization date of July 21st. Green contour indicates the border line of Pakistan. The saliency maps (c) were generated using the gradient of the negative standardized $\textrm{TP}$ anomaly, averaged over the Pakistan region, in relation to the input $\textrm{SST}$. These maps correspond to forecasts initialized on July 28th (07-28, first column) and July 21st (07-21, second column). Here, the red and blue colors indicate the positive and negative correlations between the negative of standardized $\textrm{TP}$ and variations in $\textrm{SST}$. The black lines on the bars in Figure 4 represent the 25th and 75th percentiles of the ensemble forecasts for each start date for both ECMWF and FuXi-S2S models.}
    \label{event}    
\end{figure}

Figure \ref{event} illustrates the observed standardized $\textrm{TP}$ anomaly alongside predictions that were initialized on different dates, generated by both the FuXi-S2S and ECMWF S2S models. These observations, taken from the Global Precipitation Climatology Project (GPCP), are spatially averaged over the Pakistan region (60 to 70\textdegree E in longitude and 25 to 35\textdegree N in latitude), and temporally over a two-week period from August 16th to August 31st, 2022, corresponding to the period of most intense rainfall. The standardized anomaly for observed rainfall is approximately 6 standard deviations above the climatological mean. It is evident that the ECMWF S2S model considerably underestimates rainfall intensity for forecasts initialized on July 21st, achieving only about one-third of the observed values. The ECMWF S2S forecasts gradually converge toward observations as the initialization dates approach the actual event. In contrast, FuXi-S2S exhibits superior forecast performance in predicting the intensity of extreme rainfall events earlier compared to ECMWF S2S. Specifically, FuXi-S2S predicts rainfall levels of at least 4 standard deviation above the climatological mean for forecasts initialized on July 21st, which is approximately 4 weeks in advance. Moreover, the spatial distributions of the standardized $\textrm{TP}$ anomaly reveal that the FuXi-S2S predicted $\textrm{TP}$ pattern more closely matches the observations.

Forecast skill typically improves with decreasing lead time, as in the ECMWF S2S model. The rainfall anomaly grows in FuXi-S2S forecasts initialized on July 28 (lead time of 18 days), albeit with a large forecast spread, possible due to SST influence. Indeed, the saliency maps show that the FuXi-S2S forecasts initialized on July 28 and July 21 successfully captured predictabable signals from SST anomalies in the tropical central Pacific and western Indian Ocean (Figures \ref{event}c). At shorter lead times, the SST influence decreases while the effect of atmospheric initial conditions increases. The varying importance of SST and initial conditions may cause variability in the FuXi-S2S forecasts with lead time.

\subsection{Discovery of precursor signals for the 2022 Pakistan floods prediction}

Data-driven machine learning forecasting models, such as FuXi-S2S, often lack explicit integration of prior knowledge about the physical system they aim to predict. As a result, they are often referred to as 'black boxes'. Although FuXi-S2S has shown accuracy in previous subsections, the opacity of its predictive processes can diminish confidence in its reliability. Therefore, it is imperative to interpret FuXi-S2S, ensuring that their underlying reasoning is consistent with established understanding of weather systems. Here, we generated saliency maps to disentangle the key driving processes behind the FuXi-S2S model's prediction of the 2022 floods in Pakistan. 

In this study, we utilized the negative absolute values of the $\textrm{TP}$ anomaly, averaged across the Pakistan region (outlined by the green box in Figure \ref{event}c), as a loss function. By implementing backward propagation of this loss function to calculate gradients, we obtained the saliency maps. These maps use red and blue colors to signify positive and negative correlations, respectively between the negative of standardized $\textrm{TP}$ anomaly and $\textrm{SST}$. Specifically, blue (red) areas indicate that a decrease (increase) in $\textrm{SST}$ is associated with an increase (decrease) in the negative of standardized $\textrm{TP}$ anomaly, thereby leading to an increase (decrease) in $\textrm{TP}$ anomaly. Analysis of these saliency maps facilitated the identification of potential precursor signals and sources of predictability that contributed to the occurrence of the extreme $\textrm{TP}$ event. As illustrated in Figure \ref{event}c, $\textrm{SST}$ precursor signals, identified in forecasts initialized on different dates (July 28th and July 21st in 2022), show remarkable consistency. These signals indicate a consistent cooling of $\textrm{SST}$ in the equatorial central Pacific and the tropical western Indian Ocean, along with warming in the tropical eastern Pacific. This spatial pattern aligns closely with findings from previous studies \cite{dunstone2023windows}, which pinpointed the rapid development of a La Niña in the tropical Pacific and a negative phase of the Indian Ocean Dipole (IOD) in the summer of 2022 as key precursor signals and driving forces of Pakistan's intense $\textrm{TP}$ event. Our results confirm that the high predictive skill of the FuXi-S2S model can be attributed to its effective capture of the primary predictable sources of this event. Furthermore, these findings demonstrate the model's potential as a valuable tool for rapidly exploring the mechanisms behind extreme events and uncovering teleconnections within Earth's systems, thereby enhancing our physical understanding. Here, we focus on the gradient with respect to $\textrm{SST}$. Nevertheless, it is important to acknowledge the existence of other significant precursor signals that may be associated with this extreme event, including $\textrm{U}$, $\textrm{V}$, and $\textrm{Z}$ anomalies as noted in \cite{hong2023causes}. A more comprehensive examination of these factors is intended for future research.

\section{Discussion} 

In this paper, we introduced FuXi-S2S, a machine learning based subseasonal forecasting model. This model provides global forecasts of daily mean values for up to 42 days, with a daily temporal resolution and 1.5$^{\circ}$ spatial resolution encompassing five upper-air atmospheric variables across 13 pressure levels and 11 surface variables. The performance of FuXi-S2S was rigorously evaluated against ERA5 reanalysis data and compared with ECMWF S2S reforecasts. A comprehensive suite of metrics was employed for this evaluation, including the deterministic metrics of the ensemble mean, the probabilistic metrics of the ensemble forecast, and the capability to predict extreme events. Our results demonstrated that FuXi-S2S surpasses ECMWF S2S in forecast accuracy for the evaluated variables. Furthermore, FuXi-S2S significantly improves accuracy in predicting the MJO, extending the skillful MJO prediction from 30 days to 36 days. This improvement is particularly important given the MJO's influence on global climate patterns, and consequently, it improves the model's ${\textrm{TP}}$) forecast accuracy globally. Moreover, FuXi-S2S has shown utility in practical scenarios, such as its superior performance in predicting the extreme rainfall during the 2022 Pakistan floods earlier than the ECMWF S2S model. This early prediction capability is vital for improving disaster preparedness and response. 

A key contributor to the superiority of FuXi-S2S is its innovative method of generating perturbations, which is essential for its successful ensemble forecasting. Unlike conventional models that employ random or meticulously calculated perturbations in initial conditions, FuXi-S2S incorporates background flow-dependent perturbations into its hidden features. These flow-dependent perturbations have shown to significantly enhance model's subseasonal forecast performance, as illustrated in Supplementary Figure \ref{flow_dependent}.
%速度快我们可以快速产生很多member
FuXi-S2S, as a machine learning model, also distinguishes itself by its ability to generate large ensembles forecasts rapidly and efficiently, requiring significantly less time and computational resources than traditional models. Specifically, it can complete a comprehensive 42-day forecast with daily time steps in approximately 7 seconds using an Nvidia A100 GPU for a single member. Ensemble size is a critical determinant of the ensemble forecast skill. Research suggests that the optimal number of members for subseasonal forecasts potentially falls within the range of 100 to 200 members \cite{Buizza2019}. To ensure a fair comparison with the ECMWF S2S model, we have currently limited the FuXi-S2S model to a 51-member ensemble. However, it's important to note that FuXi-S2S is capable of generating larger ensembles with only a moderate increase in computational demands. Our supplementary Figure \ref{larger_member} illustrates that increasing the ensemble size to 101 members further enhances the forecast performance of FuXi-S2S compared to the 51-member ensemble.

%这段话有点言过其实
Beyond its computational efficiency and superior accuracy, FuXi-S2S notably excels in identifying precursor signals and disentangling the complex processes underlying climate extremes, as demonstrated by its accurate prediction of the 2022 floods in Pakistan. Many subseasonal forecasting challenges stem from the limited understanding of these complex processes. Traditional physics-based models often rely on oversimplified representations of physical processes, which diminishes their forecast performance and analytical depth. In contrast, FuXi-S2S demonstrates proficiency in learning complex patterns and identifying subtle teleconnections from vast amounts of data. This approach resonates with Albert Einstein's insight, 'You can't solve a problem with the ways of thinking that created it.'. In our study of the 2022 extreme rainfall event in Pakistan, we demonstrate that backward propagation and the resulting saliency maps successfully reveal that FuXi-S2S makes accurate forecasts by effectively capturing the key predictable sources associated with this event. Moreover, such gradient-based interpretation methods aid in explaining weather and climate forecasts made by machine learning models, such as the FuXi-S2S model \cite{yang2024interpretable}. Therefore, we advocate for a paradigm shift in the application of machine learning models like FuXi-S2S. The focus should not extend beyond enhancing forecast accuracy to include the development of a comprehensive framework for discovering previously unknown processes within the Earth's system \cite{faghmous2014big,karpatne2017theory}. We foresee a growing reliance on machine learning models like FuXi-S2S within the scientific community, acknowledging their essential role in advancing scientific discovery in Earth system science.

While FuXi-S2S offers a computationally efficient and accurate alternative to conventional NWP models for subseasonal forecasting, it also presents significant opportunities for improvement. For instance, the ECWMF S2S model runs at a spatial resolution of 36 km \cite{haiden2018evaluation}, which is considerably finer than the 1.5\textdegree resolution of FuXi-S2S. Currently, FuXi S2S predicts daily mean values up to 50 hPa and lacks critical weather parameters such as daily maximum and minimum temperatures, which are essential for some applications. Furthermore, given the known discrepancies between the ERA5 $\textrm{TP}$ data and actual observations, as noted in \cite{nogueira2020inter,Lavers2022}, GPCP observations have been utilized to evaluate the $\textrm{TP}$ forecast performance for both ECMWF S2S and FuXi-S2S (refer to Supplementary Figure \ref{gpcp_tp}). Anticipated future enhancements to the FuXi-S2S model include increasing the spatial resolution from 1.5\textdegree to 0.25\textdegree, incorporating additional weather parameters, extending the forecast beyond the current upper limit of 50 hPa, and employing more accurate $\textrm{TP}$ data sources to enhance forecast accuracy.

\section{Methods} 
 
\subsection{Data}

ERA5 stands as the fifth iteration of the ECMWF reanalysis dataset, offering a rich array of surface and upper-air variables. It operates at a remarkable temporal resolution of 1 hour and a horizontal resolution of approximately 31 km, covering data from January 1950 to the present day \cite{hersbach2020era5}. Recognized for its expansive temporal and spatial coverage coupled with exceptional accuracy, ERA5 stands as the most comprehensive and precise reanalysis archive globally. In our study, we utilize daily statistics derived from the 1-hourly ERA5 dataset, which has a spatial resolution of $1.5^\circ$ (comprising $121\times240$ latitude-longitude grid points) and a temporal resolution of 1 day. It serves as the sole data source for training the FuXi-S2S model.

Evaluating MJO predictions against MJO indices derived from satellite observed OLR data is a common practice. Therefore, alongside the ERA5 reanalysis data, a newly developed OLR dataset called the Climate Prediction Center (CPC) OLR (CBO) has emerged. Spanning from 1991 to the present day, this dataset undergoes near real-time updates. While showing slight differences in magnitude compared to the U.S. National Oceanic and Atmospheric Administration (NOAA) Advanced Very High-Resolution Radiometer (AVHRR) OLR, the CBO dataset notably exhibits a high level of similarity in both pattern and magnitude of anomalies. In our research, we utilize the CBO data, which has a spatial resolution of 1\textdegree and a temporal resolution of 1 day. This data serves as the ground truth for OLR in the identification and verification of MJO events. Furthermore, for the assessment of rainfall in the Pakistan region, observed rainfall data are sourced from the GPCP dataset \cite{adler2018global}. It is noteworthy that the MJO indices derived from ERA5 OLR data closely align with those derived from CBO OLR data.

The FuXi-S2S model forecasts a total of 76 variables, encompassing 5 upper-air atmospheric variables across 13 pressure levels (50, 100, 150, 200, 250, 300, 400, 500, 600, 700, 850, 925, and 1000 hPa), and 11 surface variables. Among the upper-air atmospheric variables are geopotential (${\textrm{Z}}$), temperature (${\textrm{T}}$), u component of wind (${\textrm{U}}$), v component of wind (${\textrm{V}}$), and specific humidity (${\textrm{Q}}$). The surface variables include 2-meter temperature (${\textrm{T2M}}$), 2-meter dewpoint temperature (${\textrm{D2M}}$), sea surface temperature (${\textrm{SST}}$), outgoing longwave radiation (${\textrm{OLR}}$), 10-meter u wind component (${\textrm{U10}}$), 10-meter v wind component (${\textrm{V10}}$), 100-meter u wind component (${\textrm{U100}}$), 100-meter v wind component (${\textrm{V100}}$), mean sea-level pressure (${\textrm{MSL}}$), total column water vapor (${\textrm{TCWV}}$), and ${\textrm{TP}}$. ${\textrm{OLR}}$ is known as the negative of top net thermal radiation (${\textrm{TTR}}$) in ECMWF convention. Table \ref{glossary} provides a comprehensive list of these variables along with their abbreviations. Variables such as ${\textrm{U100}}$ and ${\textrm{V100}}$ were selected for their potential utility in wind energy forecasting. The selection of the ${\textrm{SST}}$ is based on prior research, which suggests that slowly evolving variables like ${\textrm{SST}}$ are crucial for identifying predictable signals \cite{albers2021subseasonal,yan2021subseasonal,richter2024quantifying}. ${\textrm{OLR}}$ was selected due to its significance in representing MJO events through ${\textrm{OLR}}$ anomalies.

The model's training relies on 67 years of data spanning from 1950 to 2016, while evaluation involves a 5-year dataset from 2017 to 2021. The z-score normalization technique is employed to normalize all input and output variables, thereby ensuring uniformity in their mean and variance. For upper-air variables, the mean and standard deviation are calculated separately for different vertical levels, using only the training dataset. Additionally, the dataset for the year 2022 undergoes evaluation and comparison against the ECMWF real-time S2S forecasts, specifically concerning the catastrophic flooding in Pakistan. More detailed evaluations of ${\textrm{TP}}$ and MJO predictions for the year 2022 can be found in the supplementary material.
%The validation set contains data from year 2002 to 2016, which is used for calculating climatology of forecast data.

In certain cases, subseasonal forecasts receives regular updates through the implementation of the latest model, incorporating research discoveries tailored for operational use \cite{stan2022advances}. For instance, the ECMWF S2S reforecasts, often termed hindcasts, which are generated on-the-fly by employing the most recent model version available at the time of forecast generation. In our research, we utilize the ECMWF S2S reforecasts generated from model cycle C47r3. These reforecasts encompass initialization dates over 20 years, ranging from January 3, 2002, to December 29, 2021. The ECMWF S2S reforecasts are initialized twice weekly, aligning with the real-time forecasts. Additionally, our comparative analysis involves employing the 51-member ECMWF real-time S2S forecast for the year 2022. For the analysis using testing data from 2017 to 2021, anomalies for all variables are defined as deviations from the climatological mean calculated over the 15-year period from 2002 to 2016. Meanwhile, for the analysis based on testing data in the year 2022, the climatological mean is calculated over the period from 2002 to 2021. Furthermore, a set of hindcasts from 2002 to 2016 is generated for FuXi-S2S, which are used to to establish a climatology. This climatology is then subtracted from the FuXi-S2S forecasts for the testing data spanning from 2017 to 2021. This process facilitates the calculation of FuXi-S2S anomalies for evaluations.

To ensure equitable comparisons, we evaluate FuXi-S2S forecasts specifically on identical initialization dates corresponding to those utilized for both the ECMWF S2S reforecasts and forecasts. This approach facilitates a fair and direct assessment between FuXi-S2S and ECMWF S2S.

\subsection{FuXi-S2S model}

Most state-of-the-art machine learning models utilized in medium-range weather forecasting are built upon encoder-decoder \cite{cho2014properties} architectures \cite{bi2022panguweather,lam2022graphcast,chen2023fuxi,olivetti2023advances}. These structures are favored due to their proficiency in processing and generating sequential and spatial data. Within these architectures, the encoder processes key features from the input data, and transforms them into a compressed and abstract representation in the latent space. The decoder then utilizes this representation to generate weather forecasts. The primary objective of training these models is to minimize differences between the model's output and the target data. However, the standard encoder-decoder structures are inherently deterministic, producing identical forecasts for the same inputs, which limits their applicability in generating ensemble forecasts. To overcome this limitation, we introduce the FuXi-S2S model, drawing inspiration from Variational Autoencoders (VAEs) \cite{doersch2016tutorial,zhao2017learning,kingma2019introduction}. VAEs are inherently probabilistic, making them well-suited for tasks that require uncertainty quantification. Like VAEs, the FuXi-S2S model's encoder does not merely generate a static hidden feature from input data. Instead, it transforms input data into a Gaussian distribution in the latent space, which captures the probabilistic characteristics of the data, along with a static hidden feature. Then, the decoder combines samples from the Gaussian distribution with the static hidden feature to generate forecasts. This methodology effectively captures the inherent uncertainty in the data, thereby enabling the generation of ensemble predictions under identical input conditions by repeatedly sampling from the Gaussian distribution. For better understanding, we draw analogies between these machine learning techniques and the conventional terminology in ensemble weather/subseasonal forecasting. In our model, the static hidden feature forms the basis for deterministic forecasts, while sampling from the Gaussian distribution serves as a perturbation module. This module introduces flow-dependent perturbations into the model's hidden feature, facilitating the generation of ensemble forecasts. 

%Given the complexities of subseasonal forecasting, we use a CVAE conditioned on additional information, in contrast to a standard VAE which relies solely on the learned distribution. 
%we develop a refined Conditional Variational Autoencoder (CVAE) , tailored for subseasonal ensemble forecasting. The CVAE's encoder converts input data into a Gaussian distribution in the latent space, along with a fixed hidden feature. 

\begin{figure}[h]
    \centering
    \includegraphics[width=\linewidth]{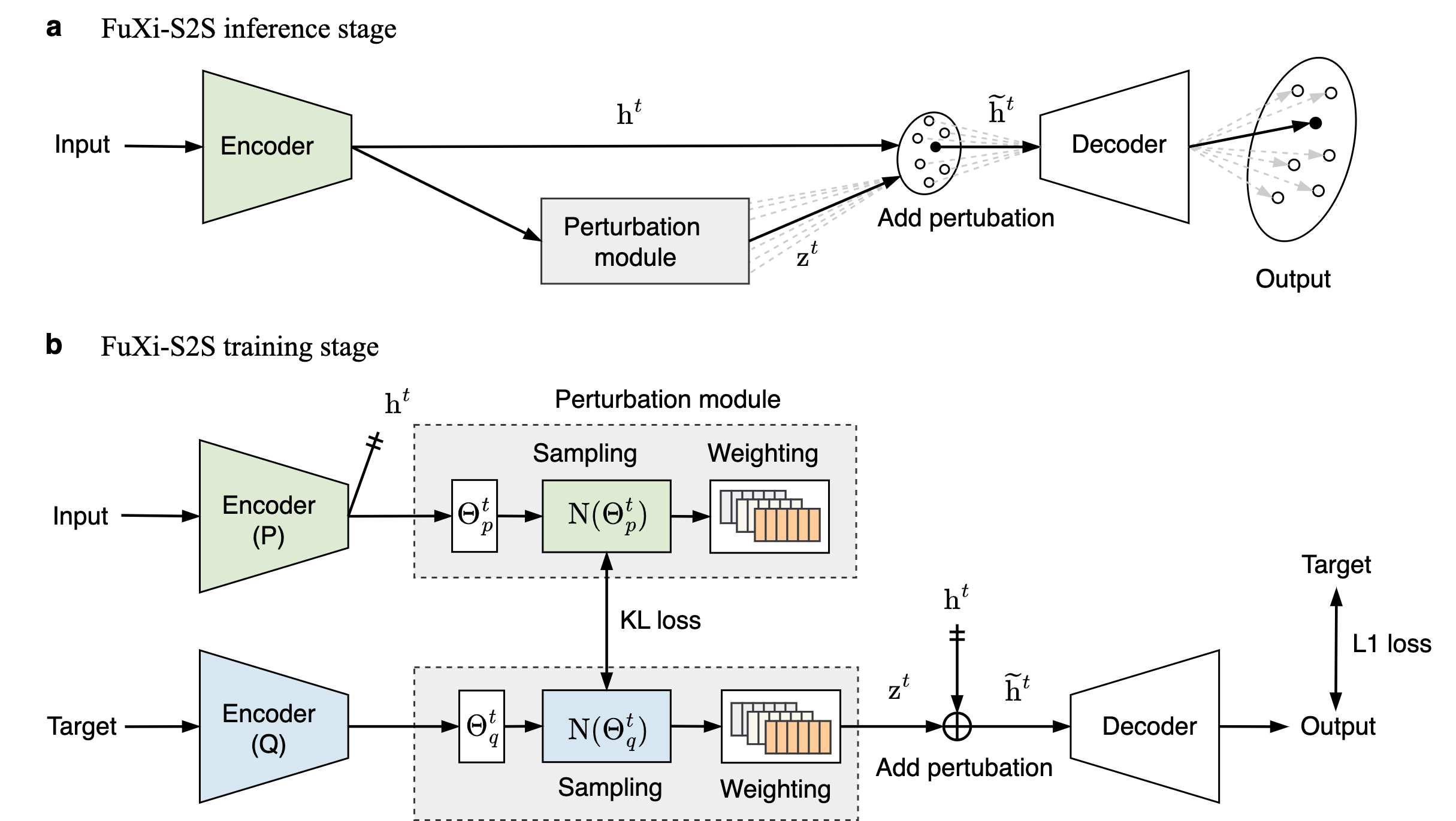}
    \caption{Schematic diagram of the structures of the FuXi Subseasonal-to-Seasonal (FuXi-S2S) model. $\textbf{a}$) Inference stage of the FuXi-S2S model. $\textrm{h}^t$ represents the hidden feature generated by the Encoder from the input data. The perturbation vector $\textrm{z}^t$ is generated by the perturbation module, resulting in the perturbed hidden feature  $\tilde{\textrm{h}}^t$. $\textbf{b}$) Training stage of the FuXi-S2S model. ${\textrm{N($\Theta$}^t}_p)$ and ${\textrm{N($\Theta$}^t}_q)$ are the low-rank multivariate Gaussian distributions generated by encoders $\textrm{P}$ and $\textrm{Q}$, respectively. The Kullback–Leibler (KL) divergence loss measures the discrepancy between the distributions predicted by both encoders, ${\textrm{N($\Theta$}^t}_p)$ and ${\textrm{N($\Theta$}^t}_q)$.}
    \label{model}    
\end{figure}

The FuXi-S2S model, illustrated in Figure \ref{model}a, consists of three primary components: an encoder $\textrm{P}$, a perturbation module, and a decoder. The encoder, processing predicted weather parameters from two preceding time steps, with each time step representing one day, as FuXi-S2S is designed to forecast daily mean values. Specifically, it takes $\hat{\textrm{\textbf{X}}}^{t-1}$ and $\hat{\textrm{\textbf{X}}}^t$ as inputs into a two-dimensional (2D) convolution layer with a kernel size of 2, which reduces the dimensions of the input data by half. Following this, the hidden feature $\textrm{h}^t$ (with dimensions of $1536\times60\times120$) is derived from 12 repeated transformer blocks. The input to the encoder is a data cube that combines both upper-air and surface variables, with dimensions of $2\times76\times121\times240$. These dimensions represent two preceding time steps (${t-1}$ and ${t}$), the number of input variables, and the latitude (${\textrm{H}}$) and longitude (${\textrm{W}}$) grid points, respectively. To account for the accumulation of forecast error over time, the forecast lead time (${t}$) is also included in the encoder's input. Besides $\textrm{h}^t$, the encoder also generates a low-rank multivariate Gaussian distribution, ${\textrm{N($\Theta$}^t}_p)$, characterized by a mean vector $\mu^t$ ($128\times60\times120$), a covariance matrix $\sigma^t$ ($1536\times60\times120$), and a diagonal covariance matrix ${diag}^t$ ($128\times60\times120$). Intermediate perturbation vectors ($\textrm{z}^t_p$, dimension: $128\times60\times120$) are sampled from this Gaussian distribution (${\textrm{N($\Theta$}^t}_p)$). These vectors, after being weighted by a learned weight vector, yield the final perturbation vectors $\textrm{z}^t$ (dimension: $1536\times60\times120$). The decoder then processes the perturbed hidden features ($\tilde{\textrm{h}}^t = \textrm{h}^t + \textrm{z}^t$) through 24 transformer blocks and a fully connected layer, resulting in the final ensemble output $\hat{\textrm{\textbf{X}}}^{t+1}$. The number of ensemble members generated equals the number of samples drawn from the Gaussian distribution ${\textrm{N($\Theta$}^t}_p)$.

The FuXi-S2S model's training primarily focuses on constructing a Gaussian distribution that accurately represents the uncertainty in the model's predictions. A significant challenge in this process is the deviation of the Gaussian distribution derived from the model's predictions from the Gaussian distribution based on the target data, largely attributable to prediction errors. This challenge is addressed through knowledge distillation, which enables the transfer of information from real-world distributions to those predicted by the model. Within this framework, the encoder $\textrm{Q}$ plays a crucial role, converting the target data into a Gaussian distribution. This distribution serves as a supervisor for the distribution generated by the encoder $\textrm{P}$, aiming to align both distributions closely by minimizing the Kullback–Leibler (KL) divergence loss ($\textrm{L}_{\textrm{KL}}$). This KL loss measures the discrepancy between the distributions predicted by both encoders. As illustrated in the Figure \ref{model}b, during the training phase of the FuXi-S2S model, the encoder $\textrm{Q}$, which shares the network structure with the encoder $\textrm{P}$, processes a data cube containing target weather parameters from a preceding and the current time steps: $\textrm{\textbf{X}}^t$ and $\textrm{\textbf{X}}^{t+1}$. It predicts a low-rank multivariate Gaussian distribution (${\textrm{N($\Theta$}^t}_q)$) similar to the encoder $\textrm{P}$. Intermediate perturbation vectors are sampled from the encoder $\textrm{Q}$'s distribution (${\textrm{N($\Theta$}^t}_q)$) during training (see Figure \ref{model}b), and from the encoder $\textrm{P}$'s distribution (${\textrm{N($\Theta$}^t}_p)$) during testing (see Figure \ref{model}a). These vectors have dimensions of $128\times60\times120$. Additionally, a L1 loss is computed between the model's output ( $\hat{\textrm{\textbf{X}}}^{t+1}$) and the target ${\textrm{\textbf{X}}}^{t+1}$. Therefore, the overall loss function at each autoregressive step is thus determined by the following equation:
\begin{equation}
    \textrm{L} = \lambda\textrm{L}_{\textrm{KL}}(\textrm{P}^t, \textrm{Q}^t) + \vert \hat{\textrm{\textbf{X}}}^{t+1} - \textrm{\textbf{X}}^{t+1}\vert
\end{equation}
where $\lambda$, a tune-able coefficient balancing $\textrm{L}_{KL}$ and $\textrm{L1}$, is set to $1\times10^{-4}$ in this study. The design of this loss function serves two purposes: the first term ensures the perturbation vector closely approximates the true data distribution, while the second term ensures the prediction unaffected by any perturbation vectors $\textrm{z}^t$.
 
%The stochastic event is unknown during each transition, thus the true prior distribution over zt is also unknown, which results in the log-likelihood of p(xt+1|x1:t, zt) to be intractable. A typical solution is to use an encoder to infer the posterior p(zt|x1:t+1) from the data x1:t+1, and then a decoder to predict xt+1 ∼pθ(xt+1|x1:t, zt) (Denton & Fergus, 2018; Kingma & Welling, 2013).
 
In this study, we employ 51 ensemble members for subseasonal ensemble forecasting. As illustrated in Supplementary Figure \ref{larger_member}, the FuXi-S2S model, when enhanced with flow-dependent perturbations incorporated into its hidden features, demonstrates considerably improved forecast performance compared to the FuXi-S2S model that combines Perlin noise in the initial conditions with fixed perturbations added to the hidden features. Notably, the addition of Perlin noise results in only marginal improvements in forecast accuracy when the ensemble size is small. However, with larger ensemble sizes, such as the 51 members in this study, the addition of Perlin noise does not enhance forecast accuracy.

%Perlin noise is also introduced into the input data. 

% Each member incorporates 3 octaves of Perlin noise with a scaling factor of 0.5, and the number of periods of noise along each axis (channel, latitude, and longitude) is set to 1, 6 and 6, respectively.

Similar to FuXi, we utilize an autoregressive, multi-step loss function to mitigate cumulative errors over long lead times, as outlined in Lam et al. \cite{lam2022graphcast}. The training process follows an autoregressive training regime and a curriculum training schedule, incrementally increasing the number of autoregressive steps from 1 to 17. Each autoregressive step undergoes 1000 gradient descent updates, resulting in a total number of 17,000 training steps. The training process utilizes 8 Nvidia A100 graphics processing units (GPUs), each employing a batch size of 1. Optimization is performed using the AdamW \cite{kingma2017adam,loshchilov2017decoupled} optimizer with the following parameters: ${\beta_{1}}$=0.9 and ${\beta_{2}}$=0.95, an initial learning rate of 2.5$\times$10$^{-4}$, and a weight decay coefficient of 0.1. The optimisation hyperparameters used for training are summarised in Supplementary Table \ref{hyper}.

%The training process takes 18000 iterations and is performed on a cluster equipped with 8 Nvidia A100 GPUs. A batch size of 1 is used on each GPU. Optimization is performed using the AdamW \cite{kingma2017adam,loshchilov2017decoupled} optimizer with the following parameters: ${\beta_{1}}$=0.9 and ${\beta_{2}}$=0.95, an initial learning rate of 2.5$\times$10$^{-5}$, and a weight decay coefficient of 0.1. To mitigate over-fitting, Scheduled DropPath \cite{larsson2017fractalnet} is applied with a dropping ratio of 0.2. 

%Figure \ref{spatial_pert} demonstrate snapshots of perturbations generated by the perturbation module in the FuXi-S2S model for the forecast initialized at 12 UTC on June 5th, 2018.

\subsection{Saliency map}

Recent developments in the field of XML have led to the emergence of various techniques \cite{samek2017explainable}, including saliency mapping. Saliency mapping quantifies the influence of a model's input on its output \cite{simonyan2013deep}. This method is characterized by the gradient intensities within the saliency maps; areas with higher gradients are considered critical by the model for making accurate predictions.

The generation of saliency maps primarily depends on backward propagation. This differs from standard model training as the propagation target can be adjusted depending on the specific goal of the analysis. Here, the saliency of the predicted anomaly relative to the input data is given by:
% \begin{equation}
% \label{saliency_equation}
%     \textrm{S} = \pdv{- \sum_{i,j \in A} \vert f^{n}(\textbf{X})_c(i,j) - \textbf{M}_c(i,j) \vert }{\textbf{X}}
% \end{equation}

\begin{equation}
\label{cost_equation}
    \textrm{J}(\textbf{X}(c_{o})) = - \sum_{i,j \in \textrm{D}} \frac{\vert f^{n}(\textbf{X})(c_{o},i,j) - \mu(c_{o},i,j) \vert}{\sigma(c_{o},i,j)}
\end{equation}
\begin{equation}
\label{saliency_equation}
    \textrm{S}(c_{i}\vert c_{o}) = \pdv{ \textrm{J}(\textbf{X}(c_{o}))}{\textbf{X}(c_{i})}
\end{equation}
where $f$ denotes the FuXi-S2S model and $n$ is the number of forward steps, while $\mu$ and $\sigma$ are the climatological mean and standard deviation, respectively. $\textrm{D}$ specify the geographical area of interest. $c_{i}$ and $c_{o}$ represent the input and output variables. A well-trained model is expected to yield a saliency map that aligns well with the established physical understanding of weather systems. In our study, we construct a aggregated saliency map by averaging the individual maps generated from each of the 51 ensemble members.

\subsection{Evaluation method}

Prior to evaluation, each variable in the 42-day forecasts undergoes a detrending process to eliminate the linear trend. This step is essential for removing the linear long-term trends potentially affected by global warming \cite{weirich2023subseasonal}. For detrending, a linear regression model is fitted to estimate the weekly mean linear trend from both forecasts and observations over the hindcast period (2002-2016). For the testing period (2017-2021), this model takes the week of the year as input data to calculate the trend, which is then subtracted from both the forecasts and observations to obtain the detrended fields. Subsequently, the deterministic metrics of the ensemble mean is evaluated using the latitude-weighted $\textrm{TCC}$, which is calculated as follows:

\begin{equation}
\label{TCC_equation}
    \textrm{TCC}(c, \tau, i, j) = \frac{\sum_{t_0 \in D} \hat{\textbf{A}}^{t_0 +\tau}_{c,i,j} \textbf{A}^{t_0 +\tau}_{c,i,j}} {\sqrt{ \sum_{t_0 \in D} (\hat{\textbf{A}}^{t_0 +\tau}_{c,i,j})^2 \sum_{t_0 \in D} (\textbf{A}^{t_0 +\tau}_{c,i,j})^2}}
\end{equation}
where ${t_0}$ represents the forecast initialization time in the testing dataset ${D}$. $H$, and $W$ denote the number of grid points in the latitude and longitude directions. The indices ${c}$, ${i}$, and ${j}$ correspond to variables, latitude and longitude coordinates, respectively. ${\tau}$ refers to the forecast lead time steps added to ${t_0}$. $\hat{\textbf{A}}^{t_0 +\tau}_{c,i,j}$ and $\textbf{A}^{t_0 +\tau}_{c,i,j}$ are the differences between the forecast or observation and the climatological mean, with the climatological mean derived from data spanning the years from 2002 and 2016.

%$\hat{\textbf{X}}^{t_0 +\tau}_{c,i,j}$ and $\textbf{X}^{t_0 +\tau}_{c,i,j}$ indicate the forecasted and observed value of variable $c$ at the time $t_0 +\tau$ for the geographical coordinate ($i, j$).

To evaluate the ensemble forecast performance, we use the $\textrm{RPSS}$ \cite{epstein1969scoring,Wilks2011} which quantifies the comparison between the cumulative squared probability errors of a given forecast and a climatological forecast. The calculation of the $\textrm{RPSS}$ metric necessitates prior determination of the ranked probability scores ($\textrm{RPS}$) for both the forecast ($\textrm{RPS}_\textrm{forecast}$) and the climatological forecast ($\textrm{RPS}_\textrm{clim}$) should be calculated first. The $\textrm{RPS}$ aggregates the squared probability errors across $K$ ($K=3$ in this work) categories, such as tercile, arranged in ascending order. The tercile bounds are determined based on the average values over either one-week or two-week periods for each corresponding verification period. These calculations of tercile bounds are performed separately for each forecast model and observation (ERA5 data). The metric assesses the accuracy with which the probability forecast predicts the actual observation category. The $\textrm{RPS}$ score is derived from the sum of the squared differences between the cumulative categorical forecast probability and its observed counterpart, where $p_{\textrm{O}(i)} = 1$ denotes the observed category and $p_{\textrm{O}(i)} = 0$ represents other categories:
\begin{equation} 
\label{RPS_forecast}
     \textrm{RPS}_\textrm{forecast} = \sum_{k=1}^{K}(\textrm{F}_{\textrm{forecast}(k)} - \textrm{F}_{\textrm{O}(k)})
\end{equation}

\begin{equation} 
\label{RPS_climat}
     \textrm{RPS}_\textrm{clim} = \sum_{k=1}^{K}(\textrm{F}_{\textrm{clim}(k)} - \textrm{F}_{\textrm{O}(k)})
\end{equation}
where $\textrm{F}_{\textrm{forecast}(k)} = \sum_{i=1}^{k}p_{\textrm{forecast}(i)}$, $\textrm{F}_{\textrm{clim}(k)} = \sum_{i=1}^{k}p_{\textrm{clim}(i)}$, $\textrm{F}_{\textrm{O}(k)} = \sum_{i=1}^{k}p_{\textrm{O}(i)}$ represent the $k$th components of the cumulative forecast, climatological, and observational distributions, respectively. And $p_{\textrm{forecast}(i)}$, $p_{\textrm{clim}(i)}$, $p_{\textrm{O}(i)}$ correspond to the forecasted, climatological, and observed probability of the event's occurrence in category $i$ ($i \le k$). Crucially, the $\textrm{RPS}$ is affected by both the forecast probabilities attributed to the observed category and the probabilities assigned to other categories. The $\textrm{RPS}$ value varies between 0 and 1, where a lower value denotes a smaller forecast probability error, and thus a more accurate forecast. Specifically, a $\textrm{RPS}$ value of 0 indicates a perfectly accurate categorical forecast. With the $\textrm{RPS}$ values of both the forecast and the climatological forecast, the $\textrm{RPSS}$ can be determined as:

\begin{equation} 
\label{RPSS}
     \textrm{RPSS} = 1- \frac{<\textrm{RPS}_\textrm{forecast}>}{<\textrm{RPS}_\textrm{clim}>}
\end{equation}
where, the brackets $<...>$ denote the average of the $\textrm{RPS}_\textrm{forecast}$ and $\textrm{RPS}_\textrm{clim}$ values across all forecast–observation pairs. Since each forecast category is equally probable by design, the climatological forecast assumes a 33$\%$ probability of occurrence for each category. The $\textrm{RPSS}$ metric serves a comparative measure against the climatological forecast. Its value range from $-\infty$ to 1, where 1 corresponds to a perfect forecast and higher values suggest better forecast performance. A positive $\textrm{RPSS}$ value indicates superior accuracy over the climatological forecast, while a negative value suggests inferior accuracy. A value of zero suggests that the forecast has no added skill compared to the climatological forecast. 

%Notablly, the ensemble size can also affect the $\textrm{RPSS}$, with small ensemble sizes exhibiting a negative bias \cite{mason2004using,weigel2007discrete}. To address this, Müller et al. \cite{muller2005debiased} and Weigel et al. \cite{weigel2007discrete} proposed a debiased version of $\textrm{RPSS}$, formulated as follows:
%\begin{equation} 
%\label{RPSS_debiased}
%     \textrm{RPSS}_D = 1- \frac{<\textrm{RPS}_\textrm{forecsst}>}{<\textrm{RPS}_\textrm{clim}> + D_0/M}
%\end{equation}
%where $M$ is the number of ensemble members, and \(D_0 = \frac{K^2 - 1}{6K}\).

Additionally, we use the BSS\cite{Wilks2011} to evaluate the performance of extreme forecasts. The BSS, a widely used metric for assessing the quality of categorical probabilistic forecasts, can be considered as a special case of the RPSS with two forecast categories \cite{weigel2007discrete}. The BSS is computed using the following equation:
\begin{equation} 
\label{BSS}
     \textrm{BSS} = 1- \frac{<\textrm{BS}_\textrm{forecast}>}{<\textrm{BS}_\textrm{clim}>}
\end{equation}
where $\textrm{BS}_\textrm{forecast}$ and $\textrm{BS}_\textrm{clim}$ represent the Brier Scores (BS) \cite{Brier1950} for the model's forecast and the climatological forecast, respectively. Similar to the RPS, the BS quantifies the mean squared difference between the predicted probabilities and observations (either 0 or 1) in binary probabilistic forecasts. In this study, the BSS is calculated for the ensemble mean of both FuXi-S2S and ECMWF S2S, using the 90th climatological percentiles as the threshold for extreme events. The BS ranges from 0 to 1, with lower values indicating a better agreement between ensemble forecasts and observations with 0 suggesting the best possible BS score. On the contrary, a higher BSS, up to a maximum of 1, indicates better performance. The BSS measures the improvement of a forecast's BS ($\textrm{BS}_\textrm{forecast}$) relative to that of a climatological forecast ($\textrm{BS}_\textrm{clim}$) as reference. A BSS of one indicates a perfect forecast, zero denotes no improvement over climatology, and negative values suggest inferior performance compared to climatology.

The evolution of MJO is typically characterized using the Real-time Multivariate MJO (RMM) index, as originally developed by Wheeler and Hendon \cite{wheeler2004all}. The RMM1 and RMM2 indices represent the first and second principal components of the combined Empirical Orthogonal Function (EOF). This EOF is derived based on the daily mean values of OLR, zonal wind at 850 hPa ($\textrm{U850}$), and zonal wind at 200 hPa ($\textrm{U200}$), all averaged within the latitude range of 15\textdegree N and 15\textdegree S \cite{rashid2011prediction}. In this study, we use the EOFs derived by Wheeler and Hendon (2004) \cite{wheeler2004all}. To obtain the predicted MJO indices, data from both the FuXi-S2S and ECMWF S2S models are firstly interpolated from a spatial resolution of 1.5\textdegree to a 2.5\textdegree, and projected onto the observed EOFs. After calculating the ensemble mean anomalies, the RMM for the ensemble mean of both modes was derived. The amplitude and phase of the MJO are respectively defined by the formulas: \(\textrm{RMMA}=\sqrt{\textrm{RMM1}^2(t) + \textrm{RMM2}^2(t)}\) and \(\textrm{$\theta$}=tan^{-1}\frac{\textrm{RMM2}^2(t)}{\textrm{RMM1}^2(t)}\). To assess the quality of the MJO forecasts, we calculate the bivariate $\textrm{COR}$ using the following equation:
%To evaluate the MJO forecast quality, we calculate the bivariate $\textrm{COR}$ and bivariate RMSE using the following equations:
\begin{equation}
\label{cor_MJO}
     \textrm{COR}(\tau) = \frac{\sum_{t=1}^{N}[a_1(t)b_1(t,\tau) + a_2(t)b_2(t,\tau)]} { \sqrt{ \sum_{t=1}^{N}[a_1^2(t) + a_2^2(t)] }\sqrt{ \sum_{t=1}^{N}[b_1^2(t,\tau) + b_2^2(t,\tau)] } }
\end{equation}
where $a_1(t)$ and $a_2(t)$ are the observed RMM1 and RMM2 at time $t$ derived from the ERA5 reanalysis dataset. Correspondingly, $b_1(t,\tau)$ and $b_2(t,\tau)$ represent the forecasts for time $t$ with a lead time of $\tau$ days, respectively. $N$ denotes the number of total predictions. We apply the threshold of $\textrm{COR} = 0.5$ for skillful prediction \cite{rashid2011prediction}.

Additionally, we assessed the respective contributions of amplitude and phase to the prediction skills of the MJO by examining the $\textrm{COR}$ and error metrics of ensemble mean forecasts for each component. The $\textrm{COR}$ for amplitude ($\textrm{COR}_{\textrm{amplitude}}$) and phase ($\textrm{COR}_{\textrm{phase}}$) were calculated using the methods outlined by Wang et al. \cite{wang2019prediction} as follows:
\begin{equation}
\label{cor_MJO_amplitude}
     \textrm{COR}_{amplitude}(\tau) = \frac{\sum_{t=1}^{N}\textrm{RMMA}_{\textrm{obs}}(t) \times \textrm{RMMA}_{\textrm{forecast}}(t,\tau)} { \sqrt{ \sum_{t=1}^{N}\textrm{RMMA}_{\textrm{obs}}^2(t)}\sqrt{ \sum_{t=1}^{N}\textrm{RMMA}_{\textrm{forecast}}^2(t,\tau)} }
\end{equation}

\begin{equation}
\label{cor_MJO_phase}
     \textrm{COR}_{phase}(\tau) = \frac{\sum_{t=1}^{N}\textrm{RMMA}_{\textrm{obs}}(t) \times cos(\textrm{$\theta$}_{\textrm{forecast}}(t,\tau) - \textrm{$\theta$}_{\textrm{obs}}(t))} { \sum_{t=1}^{N}\textrm{RMMA}_{\textrm{obs}}^2(t) }
\end{equation}
where $\textrm{RMMA}_{\textrm{obs}}$ and $\textrm{RMMA}_{\textrm{forecast}}$ represent the observed and predicted amplitudes of the MJO, respectively, while $\textrm{$\theta$}_{\textrm{obs}}$ and $\textrm{$\theta$}_{\textrm{forecast}}$ denote the observed and predicted phases.
Additionally, we computed the average amplitude and phase errors ($\textrm{ERROR}_{\textrm{amplitude}}$ and $\textrm{ERROR}_{\textrm{phase}}$) as follows, based on the method described by Rashid et al. \cite{rashid2011prediction}:
\begin{equation}
     \textrm{ERROR}_{\textrm{amplitude}}(\tau) = \frac{1}{N}\sum_{t=1}^{N}( \textrm{RMMA}_{\textrm{forecast}}(t,\tau) - \textrm{RMMA}_{\textrm{obs}}(t) )
\end{equation}
\begin{equation}
     \textrm{ERROR}_{\textrm{phase}}(\tau) = \frac{1}{N}\sum_{t=1}^{N}tan^{-1}( \frac{a_1(t)b_2(t,\tau) - a_2(t)b_1(t,\tau)}{a_1(t)b_1(t,\tau) + a_2(t)b_2(t,\tau)} )
\end{equation}
Further details about the $\textrm{COR}$ and $\textrm{ERROR}$ for the amplitude and phase are presented in the Supplementary Figure \ref{mjo_amp_phase}.

%We apply the threshold of $\textrm{COR} = 0.5$ and $\textrm{RMSE} = 1.4$ for skillful prediction \cite{rashid2011prediction}.
%\begin{equation} 
%\label{rmse_MJO}
%     \textrm{RMSE}(\tau) = \sqrt{\frac{1}{N} \sum_{t=1}^{N}[\vert a_1(t) - b_1(t, \tau)\vert ^2 + \vert a_2(t) - b_2(t, \tau)\vert^2] }
%\end{equation}

Atmospheric predictability exhibits significant day-to-day variability, which in turn affects the potential accuracy of weather forecasts. To determine whether FuXi-S2S consistently outperform ECMWF S2S despite this variability, we adopted a bootstrapping approach for significance testing. This method involves generating a large number of synthetic datasets, for example 1000 in this work. For each day within these datasets, a forecast is randomly selected from either model A or model B. The forecast skill of each synthetic dataset is then evaluated by comparing it with actual observation. If the performance of model A surpasses the 97.5th percentile of the skill distribution derived from the synthetic datasets, it can be considered  “significantly better” than model B. In contrast, if its performance falls below the 2.5th percentile, it is regarded as “significantly worse”. We also analyzed where the FuXi-S2S and ECMWF S2S models are significantly better or worse than the climatological forecasts, with model B representing these forecasts. Throughout the paper, significance testing has been applied to all bar plots and spatial map of statistical metrics. For all the bar plots in the paper, a pale color is used when the FuXi-S2S model do not show a statistically significant improvement over the ECMWF S2S model. Additionally, we have marked areas on all spatial maps where the skill score is statistically significant with stippling.

\section*{Data Availability Statement}
We downloaded a subset of the daily statistics from the ERA5 hourly data from the official website of Copernicus Climate Data (CDS) at \url{https://cds.climate.copernicus.eu/cdsapp#!/software/app-c3s-daily-era5-statistics}. The ECMWF S2S data were obtained from \url{https://apps.ecmwf.int/datasets/data/s2s/}. The 1\textdegree CPC OLR data are provided by the NOAA Physical Sciences Laboratory (PSL) from their website of \url{https://psl.noaa.gov}. Rainfall data from the Global Precipitation Climatology Project (GPCP) was obtained from the National Oceanic and Atmospheric Administration (NOAA), specifically the National Centers for Environmental Information (NCEI), which is accessible at \url{https://www.ncei.noaa.gov/products/global-precipitation-climatology-project}.

The relevant data from each figure in the main manuscript and in the Supplementary Information are provided in \url{https://zenodo.org/records/12662702} \cite{data2024}.

%NOAA Interpolated Outgoing Longwave Radiation (OLR) were downloaded from \url{https://psl.noaa.gov/data/gridded/data.olrcdr.interp.html}

%去掉谷歌网盘链接
\section*{Code Availability Statement}
The source code employed for training and running FuXi-S2S models in this research is accessible within a specific Google Drive folder (\url{https://drive.google.com/drive/folders/1z47CRQdKFZaOjtKQWSNZobC1_RePUVIK?usp=sharing}) \cite{code2023}. As the FuXi-S2S model and code are essential resources for this study. Currently, access to these resources is limited.

Calculation of MJO index is based on the EOFs derived by Wheeler and Hendon (2004) \cite{wheeler2004all}.

The implementation of Perlin noise is based on publicly available from the GitHub repository: \url{https:// github.com/pvigier/perlin-numpy}.

%\section*{Authors' contributions}
%L.C. and H.L. designed the research. L.C. performed the model training and evaluation. L.C. and X.Z. wrote the manuscript.

\noindent


\bibliography{refs}% common bib file


\section*{Acknowledgements}
This work was supported by the National Key R\&D Program of China under Grant 2021YFA0718000, National Natural Science Foundation of China under Grant 42175052, and China Meteorological Administration (CMA) Youth Innovation Team (CMA2024QN06. We extend our sincere gratitude to the researchers at ECMWF for their invaluable contributions to the collection, archival, dissemination, and maintenance of the ERA5 reanalysis dataset and ECMWF S2S reforecast and real-time forecast data.

\section*{Author Contributions}
H.L., X.Z, L.C., and B.L. designed the project. L.C. designed and performed the model training. X.Z. and L.C. performed the analysis under supervision of H.L., B.L., W.J., Q.C., L.W., C.L., Z.H., and Y.Q.. X.Z. and L.C. wrote and revised the manuscript. J.W., D.C., and S.X. contributed to interpreting results and discussions of associated dynamics.

\section*{Competing interests}
The authors declare no competing interests.

\section*{Tables}

\begin{table}
\centering
\caption{\label{glossary} A summary of all the upper-air and surface variable names and their abbreviations in this paper.}
\begin{tabularx}{\textwidth}{cXX}
\hline
Type & Full name & Abbreviation \\
\hline
upper-air variables & geopotential & ${\textrm{Z}}$  \\
                    & temperature  & ${\textrm{T}}$  \\
                    & u component of wind & ${\textrm{U}}$ \\
                    & v component of wind & ${\textrm{V}}$ \\
                    & specific humidity & ${\textrm{Q}}$ \\
                    \hline
surface variables   & 2-meter temperature & ${\textrm{T2M}}$ \\
                    & 2-meter dewpoint temperature & ${\textrm{D2M}}$ \\
                    & sea surface temperature & ${\textrm{SST}}$ \\
                    & outgoing longwave radiation & ${\textrm{OLR}}$ \\ 
                    & 10-meter u wind component & ${\textrm{U10}}$ \\
                    & 10-meter v wind component & ${\textrm{V10}}$ \\
                    & 100-meter u wind component & ${\textrm{U100}}$ \\
                    & 100-meter v wind component & ${\textrm{V100}}$ \\
                    & mean sea-level pressure & ${\textrm{MSL}}$ \\
                    & total column water vapor & ${\textrm{TCWV}}$ \\
                    & total precipitation & ${\textrm{TP}}$ \\                    
\hline
\end{tabularx}
\end{table}

\section*{Figure Legends}

\FloatBarrier

\FloatBarrier

\FloatBarrier

\FloatBarrier

\FloatBarrier

\end{CJK*}

\clearpage

\appendix

\renewcommand\thefigure{\thesection.\arabic{figure}}    

\setcounter{figure}{0}    

\end{document}

\begin{CJK*}{UTF8}{gbsn}

%\title[Article Title]{FuXi: A foundation model for 15 days global weather forecast}
\title[FuXi-S2S]{Supplementary Information: Meet the challenge of predictability desert: a machine learning model that outperforms conventional global subseasonal forecast models}


\author[1,2]{\fnm{Lei} \sur{Chen}}\email{cltpys@163.com}
\equalcont{These authors contributed equally to this work.}
\author[1]{\fnm{Xiaohui} \sur{Zhong}}\email{x7zhong@gmail.com}
\equalcont{These authors contributed equally to this work.}

\author*[1]{\fnm{Hao} \sur{Li}}\email{lihao$\_$lh@fudan.edu.cn}
\equalcont{These authors contributed equally to this work.}

\author[3]{\fnm{Jie} \sur{Wu}}\email{wujie@cma.gov.com}
\equalcont{These authors contributed equally to this work.}

\author*[3,4]{\fnm{Bo} \sur{Lu}}\email{bolu@cma.gov.cn}

\author[5]{\fnm{Deliang} \sur{Chen}}\email{deliang@gvc.gu.se}

\author[6]{\fnm{Shang-Ping} \sur{Xie}}\email{sxie@ucsd.edu}

\author[7,8,9]{\fnm{Libo} \sur{Wu}}\email{wulibo@fudan.edu.cn}

\author[3]{\fnm{Qingchen} \sur{Chao}}\email{chaoqc@cma.gov.cn}

\author[1]{\fnm{Chensen} \sur{Lin}}\email{linchensen@fudan.edu.cn}

\author[1]{\fnm{Zixin} \sur{Hu}}\email{huzixin@fudan.edu.cn}

\author*[2,1]{\fnm{Yuan} \sur{Qi}}\email{qiyuan@fudan.edu.cn}

\affil[1]{\orgdiv{Artificial Intelligence Innovation and Incubation Institute}, \orgname{Fudan University}, \orgaddress{\city{Shanghai}, \postcode{200433}, \country{China}}}

\affil[2]{\orgname{Shanghai Academy of Artificial Intelligence for Science}, \orgaddress{\city{Shanghai}, \postcode{200232}, \country{China}}}

\affil[3]{\orgdiv{China Meteorological Administration Key Laboratory for Climate Prediction Studies}, \orgname{National Climate Center}, \orgaddress{\city{Beijing}, \postcode{100081}, \country{China}}}

\affil[4]{\orgname{Xiong'an Institute of Meteorological Artificial Intelligence}, \orgaddress{\city{Xiong'an}, \country{China}}}

\affil[5]{\orgname{University of Gothenburg}, \orgaddress{\country{Sweden}}}

\affil[6]{\orgdiv{Scripps Institution of Oceanography}, \orgname{University of California San Diego}, \orgaddress{\country{USA}}}

\affil[7]{\orgdiv{School of Data Science}, \orgname{Fudan University}, \orgaddress{\city{Shanghai}, \postcode{200433}, \country{China}}}

\affil[8]{\orgdiv{Institute for Big Data}, \orgname{Fudan University}, \orgaddress{\city{Shanghai}, \postcode{200433}, \country{China}}}

\affil[9]{\orgdiv{MOE Laboratory for National Development and Intelligent Governance}, \orgname{Fudan University}, \orgaddress{\city{Shanghai}, \postcode{200433}, \country{China}}}

\maketitle

\section*{Contents of this file}
\begin{itemize}
  \item[] Supplementary Notes 1 to 8
  \item[] Supplementary Figures 1 to 16
  \item[] Supplementary Table 1
\end{itemize}

\section*{Supplementary Notes}

\section{Effectiveness of flow-dependent perturbations}
\label{effectiveness}
This section discusses the effect of incorporating the flow-dependent perturbations into the model's hidden features to enhance performance in subseasonal forecasts. We conducted experiments using FuXi-S2S models which exclusively employ Perlin noise in the initial conditions or combine Perlin noise in the initial conditions with fixed perturbations added into the hidden features, to generate 42-day forecasts. Subsequently we evaluate their performance in comparison with the original FuXi-S2S model. 

\begin{figure}[h]
    \centering
    \includegraphics[width=\linewidth]{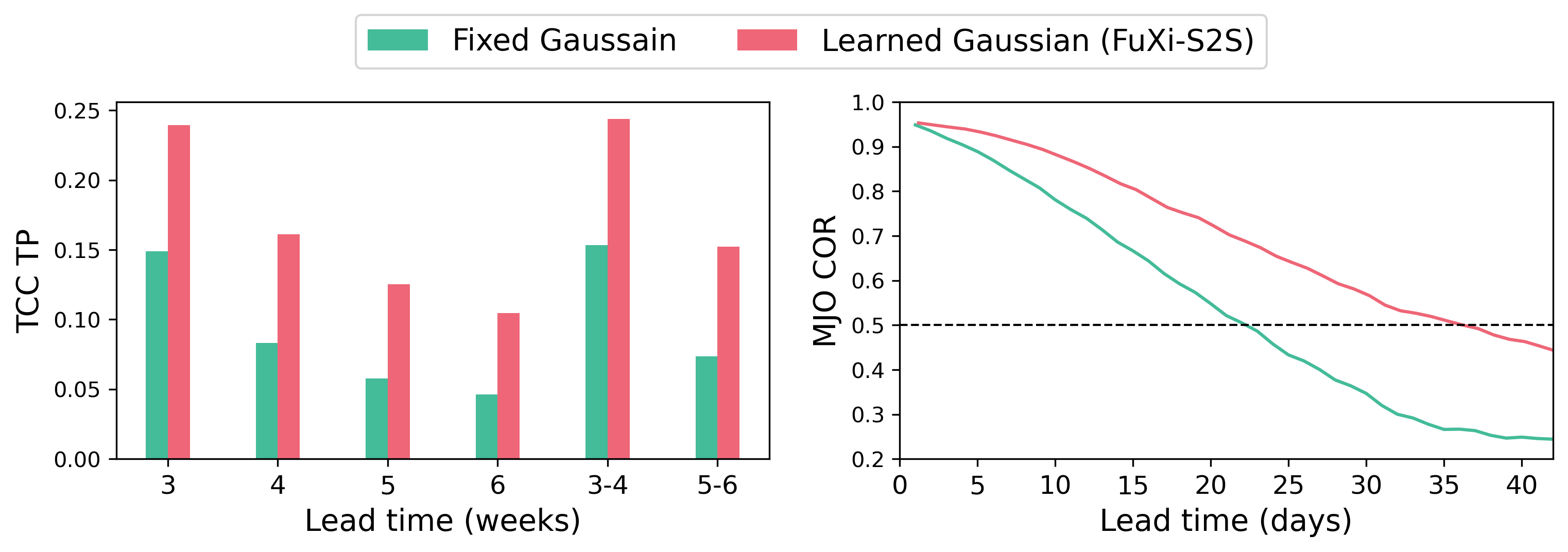}
    \caption{Comparison of the FuXi-S2S model (in red) and FuXi-S2S with fixed Gaussian perturbations (in green), utilizing all testing data from 2017 to 2021. The first column is the comparison of the globally-averaged latitude-weighted $\textrm{TCC}$. The second column is the comparison of the globally-averaged latitude-weighted RMM bivariate $\textrm{COR}$ of the FuXi-S2S (in red) and FuXi-S2S with fixed Gaussian noise (in light red) using testing data from 2017 to 2021. When the FuXi-S2S forecasts fail to show a statistically significant improvement over the ECMWF S2S reforecasts at the 97.5\% confidence level, a pale color scheme is used to denote these results.}
    \label{flow_dependent}    
\end{figure}

Supplementary Figure \ref{flow_dependent} presents a comparison of the globally-averaged and latitude-weighted ${\textrm{TCC}}$ for ${\textrm{TP}}$. This analysis encompasses all testing data from the period spanning from 2017 to 2021. The FuXi-S2S model, which incorporates flow-dependent perturbations into its hidden features, consistently exhibits considerably improved forecast performance in comparison to the FuXi-S2S model that incorporates fixed Gaussian noise into the hidden features, across all forecast lead times. Furthermore, the introduction of flow-dependent perturbations has extended the FuXi-S2S model's skillful MJO prediction from 22 days to 36 days.

\section{Deterministic forecast metrics comparison}

Supplementary Figure \ref{TCC_area_bar} presents a comparison of latitude-weighted $\textrm{TCC}$ between FuXi-S2S and ECMWF S2S. It examines $\textrm{TP}$, $\textrm{T2M}$, $\textrm{Z500}$, and $\textrm{OLR}$ across four geographical regions: in the extra-tropics (90\textdegree S - 30\textdegree S and 30\textdegree N - 90\textdegree N), in the tropics (30°S - 30°N), over land, and over the ocean. Within the extra-tropical regions, FuXi-S2S consistently exhibits superior performance compared to ECMWF S2S for all four variables. In tropical regions, FuXi-S2S outperforms ECMWF S2S for $\textrm{TP}$ and $\textrm{OLR}$, while achieving comparable accuracy in $\textrm{T2M}$ and $\textrm{Z500}$. Over land areas, FuXi-S2S demonstrates consistently higher \textrm{TCC} values for $\textrm{TP}$, $\textrm{Z500}$, and $\textrm{OLR}$.

Supplementary Figure \ref{rmse} presents a comparison of the globally-averaged and latitude-weighted root mean square error ($\textrm{RMSE}$) of the ensemble mean between ECMWF S2S real-time forecasts and FuXi-S2S forecasts for total precipitation (${\textrm{TP}}$), 2-meter temperature (${\textrm{T2M}}$), geopotential at 500 hPa (${\textrm{Z500}}$), and outgoing longwave radiation (${\textrm{OLR}}$). The analysis is derived from the averaged $\textrm{RMSE}$ computed using testing data from the year 2022. FuXi-S2S demonstrates superior forecast performance for all four variables across all forecast lead times compared to ECMWF S2S, consistently achieving lower $\textrm{RMSE}$ values than ECMWF S2S.

Supplementary Figure \ref{spectra} presents the energy spectra of $\textrm{T2M}$, $\textrm{Z500}$, $\textrm{TP}$, and $\textrm{OLR}$ at seven forecast lead times: 1, 8, 15, 22, 29, 36, and 42 days). This figure demonstrates the effectiveness of the models by showcasing the energy levels across various scales and lead times. The spectra are calculated and presented for both the ensemble mean and a randomly selected ensemble member from the ECMWF S2S reforecasts and FuXi-S2S forecasts. The ERA5 spectra remain consistent across increasing forecast lead times, serving as a baseline to evaluate whether the forecasts become increasingly smoother as the forecast lead times increases. Remarkably, at longer wavelengths, a randomly selected member from either the FuXi-S2S or ECMWF S2S models shows closer alignment with the ERA5 benchmark, suggesting that both models proficiently predict the dominant, larger-scale motions. However, at shorter wavelengths, the FuXi-S2S model initially matches the ERA5 spectra but shows a gradual reduction in energy as forecast lead times increase, indicating increasingly smoother forecasts at smaller scales.  In contrast, an ECMWF S2S ensemble member maintains consistent agreement at these smaller scales. Regarding the ensemble mean, both the ECMWF S2S reforecasts and FuXi-S2S forecasts generally exhibit lower energy spectra levels at most forecast lead times compared to both ERA5 data and individual ensemble members. As the lead time increases, the ensemble mean of both models demonstrate a decline in performance at smaller scales, a degradation more significant than that observed in individual ensemble members of the FuXi-S2S model. The notably lower energy levels in the FuXi-S2S model, particularly at longer forecast lead times compared to the ECMWF S2S and ERA5 data, underscore a critical area for model improvement to enhance forecast accuracy and smoothness.

\section{Ensemble forecast metrics comparison}

Supplementary Figure \ref{spread_skill_plot} compares the globally-averaged and latitude-weighted $\textrm{RMSE}$, ensemble spread, and spread skill ratio (${\textrm{SSR}}$) between ECMWF S2S reforecasts and FuXi-S2S forecasts for $\textrm{TP}$, $\textrm{T2M}$, $\textrm{Z500}$, and sea surface temperature (${\textrm{SST}}$). These metrics are derived from daily mean forecasts, calculated using all available testing data from 2017 to 2021 as a function of forecast lead times. For TP, FuXi-S2S consistently outperforms ECMWF S2S in terms of RMSE. For SST, FuXi-S2S initially shows slightly superior performance compared to ECMWF S2S for the forecast lead times of 15 to 20 days, but its performance declines relative to ECMWF S2S thereafter. In terms of ensemble spread, FuXi-S2S generally shows smaller spread than ECMWF S2S for both TP and SST. However, their $\textrm{SSR}$ values are consistently lower than those of ECMWF S2S across all forecast lead times, suggesting that the ensemble spread of ECMWF S2S more accurately predicts forecast skill for these variables. For $\textrm{T2M}$, FuXi-S2S demonstrates $\textrm{SSR}$ values closer to 1 compared to ECMWF S2S during the forecast lead times from 20 to 42 days, indicating a higher reliability of ensemble spread. Overall, both ECMWF S2S and FuXi-S2S have $\textrm{SSR}$ values below 1 for all 4 evaluated variables across all forecast lead times, suggesting underdispersion. This result suggests that there is still room for improvement in the FuXi-S2S to achieve $\textrm{SSR}$ closer to 1.

\section{MJO predictions}

The Madden-Julian Oscillation (MJO) stimulates several important teleconnection patterns, such as the Pacific-North American (PNA) pattern, which profoundly impacts extratropical anomalies. Therefore, accurately simulating MJO-related teleconnections is crucial for effective subseasonal forecasts. Consistent with previous findings \citesupp{Seo2017}, negative PNA-like patterns are observed when MJO convection anomalies are in Phases 4 (Supplementary Figure \ref{z500}). Notably, the FuXi-S2S model proficiently reproduces these anomalous circulation patterns, evidenced by its consistently high pearson correlation coefficient (PCC) even at extended forecast lead time (weeks 5 and 6). This model demonstrate superior PCC for MJO-associated ${\textrm{Z500}}$ patterns in FuXi-S2S compared to the ECMWF model across various lead times. As a result, the FuXi-S2S model's superior capability in MJO prediction and its accurate simulation of MJO teleconnections significantly enhance its performance in subseasonal forecasting, especially in extratropical regions.

Supplementary Figure \ref{mjo_amp_phase} presents a comparative analysis of the bivariate correlation coefficient ($\textrm{COR}$) and error ($\textrm{ERROR}$) metrics for the amplitude and phase of the MJO. These metrics are derived from the ensemble mean of ECMWF S2S reforecasts and FuXi-S2S forecasts, averaged over all the testing data from 2017 to 2021. Among them, the $\textrm{COR}$ reflects the accuracy of evolution, and $\textrm{ERROR}$ indicates the systematic bias. The analysis reveals that COR values decline with increasing forecast lead times, with a more pronounced decrease observed for the MJO phase compared to the amplitude. Throughout the 42-day forecast period, the $\textrm{COR}$ for MJO amplitude remains consistently above 0.8. The differences in amplitude $\textrm{COR}$ between the ECMWF S2S and FuXi-S2S models are negligible. In contrast, FuXi-S2S consistently outperforms ECMWF S2S in phase $\textrm{COR}$, maintaining higher values over the entire forecast duration. Specifically, the $\textrm{COR}$ for the MJO phase drops below 0.5 at 28 days for ECMWF S2S, whereas for FuXi-S2S, it remains above this threshold until 34 days. Additionally, negative error values for the MJO amplitude indicate that the amplitude is on average smaller in both the ECMWF S2S and FuXi-S2S simulations compared to ERA5 data, aligning with findings from previous studies \citesupp{Vitart2007,Vitart2010}. FuXi-S2S exhibits smaller errors than ECMWF S2S, suggesting it better maintains the amplitude of MJO events. Regarding the error in the MJO phase, both models show comparable values up to 32 days, indicating the small systematic phase speed error in both models. However, after 32 days, FuXi-S2S shows larger phase errors than ECMWF S2S.
Overall, the superior performance of FuXi-S2S in predicting MJO compared to that of ECMWF S2S is primarily due to its enhanced ability to predict the MJO phase.

\section{Extreme Meiyu in 2020}

The major rainy season of the East Asian summer monsoon, called Meiyu in China \citesupp{tao1987review}, typically starts in early June and ends in mid‐July. This brings abundant rainfall which accounts for the majority of the annual precipitation in China, Japan, and South Korea \citesupp{ding1992summer,yihui2005east}. In the summer of 2020, the Yangtze-Huaihe River valley (YHRV) experienced an exceptionally intense Meiyu rainy season characterized by an earlier onset and a delayed retreat. This season lasted for 62 days, making it one of the longest events since 1961, equalling the duration of the 2015 event \citesupp{liu2020characteristics}. The accumulated precipitation during the 2020 Meiyu season broke the historical record since 1961 and resulted in the most severe flooding in the YHRV in recent decades. By mid-July, the flooding had led to more than 140 fatalities or missing persons and economic losses of USD 11.75 billion.

Supplementary Figure \ref{meiyu} presents the comparison of the standardized $\textrm{TP}$ anomaly among the observations sourced from Global Precipitation Climatology Project (GPCP), ECMWF S2S, and FuXi-S2S, averaged across YHRV bounded by 105 to 125\textdegree E in longitude and 25 to 35\textdegree N in latitude. The GPCP are temporally averaged over a two-week period from June 30th to July 13th, 2020, which corresponds to a low skill and cold‐front rainy period as revealed by by Liu et al. \citesupp{liu2020record}. FuXi-S2S forecasts and ECMWF S2S reforecasts were initialized on different dates. Notably, the ECMWF S2S model predicts negative $\textrm{TP}$ anomalies for forecasts initialized on both June 2nd and June 6th. However, while the ECMWF S2S model starts to predict positive $\textrm{TP}$ anomalies from June 9th onwards, the model consistently underestimates rainfall intensity. In contrast, the FuXi-S2S model predicts positive anomalies for forecasts initialized as early as June 2nd, offering a lead time of 4 weeks prior to the occurrence of the event. Furthermore, the spatial distributions of the standardized $\textrm{TP}$ anomaly reveals that $\textrm{TP}$ patterns predicted by FuXi-S2S closely aligns with the observations, which is critical for flood preparedness. In summary, FuXi-S2S demonstrates superior performance in predicting the intensity of extreme rainfall events with longer lead time compared to ECMWF S2S.

\section{Comparisons against ECMWF S2S real-time forecasts}

This study also evaluates the performance of FuXi-S2S by analyzing testing data from 2022 and compare against the 51-member ECMWF S2S real-time forecasts from model cycle C47r3. The evaluation included deterministic metrics of the ensemble mean, ensemble metrics, and MJO forecasts.

Supplementary Figure \ref{tp_2022} presents a comparison of the globally-averaged and latitude-weighted ${\textrm{TCC}}$, ${\textrm{RMSE}}$, ${\textrm{RPSS}}$, and ${\textrm{BSS}}$ of the ensemble mean between the ECMWF S2S real-time forecasts and FuXi-S2S forecasts for ${\textrm{TP}}$ in 2022. Across all forecast lead times, FuXi-S2S demonstrates superior forecast performance in all metrics across compared to the ECMWF S2S real-time forecasts.

Supplementary Figure \ref{MJO_2022} presents the bivariate correlation ($\textrm{COR}$) skills of Real-time Multivariate MJO (RMM) index for the ensemble mean of ECMWF S2S real-time forecasts and FuXi-S2S forecasts, averaged over the testing data from 2022. When applying a $\textrm{COR}$ threshold of 0.5 to determine skillful MJO forecast, FuXi-S2S extends the skilful forecast lead time from 30 days to 41 days, surpassing the performance of ECMWF S2S real-time forecasts.

\section{Effectiveness of larger ensemble}

Supplementary Figure \ref{larger_member} presents a comparison of the globally-averaged and latitude-weighted ${\textrm{RPSS}}$ and ${\textrm{BSS}}$ of the ensemble mean between the ECMWF S2S reforecasts, the 51-member FuXi-S2S forecasts, and the 101-member FuXi-S2S forecasts, for $\textrm{T2M}$ and $\textrm{TP}$. This analysis encompasses all testing data spanning from 2017 to 2021. Notably, the 101-member FuXi-S2S demonstrate a significant improvement in forecast performance relative to the 51-member FuXi-S2S across all forecast lead times for both $\textrm{T2M}$ and $\textrm{TP}$. This enhancement proves that an increase in the number of ensemble members improves the prediction skills in subseasonal forecasts.

\section{Evaluation against the GPCP data}

Supplementary Figure \ref{gpcp_tp} and present a comparative analysis of the globally-averaged and latitude-weighted ${\textrm{TCC}}$ and ${\textrm{RPSS}}$ of the ensemble mean between ECMWF S2S reforecasts and FuXi-S2S forecasts for ${\textrm{TP}}$, based on testing data between 2017 and 2021. Unlike prior analyses, this evaluation employs the GPCP dataset as the reference, rather than the ERA5 dataset. Consistent with the results shown in Figure 1 of the main text and Supplementary Figure \ref{RPSS_area_bar}, where ERA5 serves as the verification target, the FuXi-S2S model generally outperforms ECMWF S2S at most forecast lead times, achieving higher $\textrm{TCC}$, ${\textrm{RPSS}}$, and ${\textrm{BSS}}$ values than ECMWF S2S. However, an exception is noted in week 3, where ECMWF S2S exhibits superior ${\textrm{RPSS}}$ values. Notably, since the FuXi-S2S is trained on ${\textrm{TP}}$ data from the ERA5 dataset, its performance slightly diminishes when evaluated against the GPCP dataset. This reduction in performance is likely due to the discrepancies between the GPCP and ERA5 datasets. Considering the known differences between the ERA5 ${\textrm{TP}}$ data and actual observations, as highlighted in \citesupp{nogueira2020inter,Lavers2022}, exploration of more accurate ${\textrm{TP}}$ data sources is planned to enhance the forecast accuracy of the FuXi-S2S model.

\section*{Supplementary Figures}

\FloatBarrier

\begin{figure}[h]
    \centering
    \includegraphics[width=\linewidth]{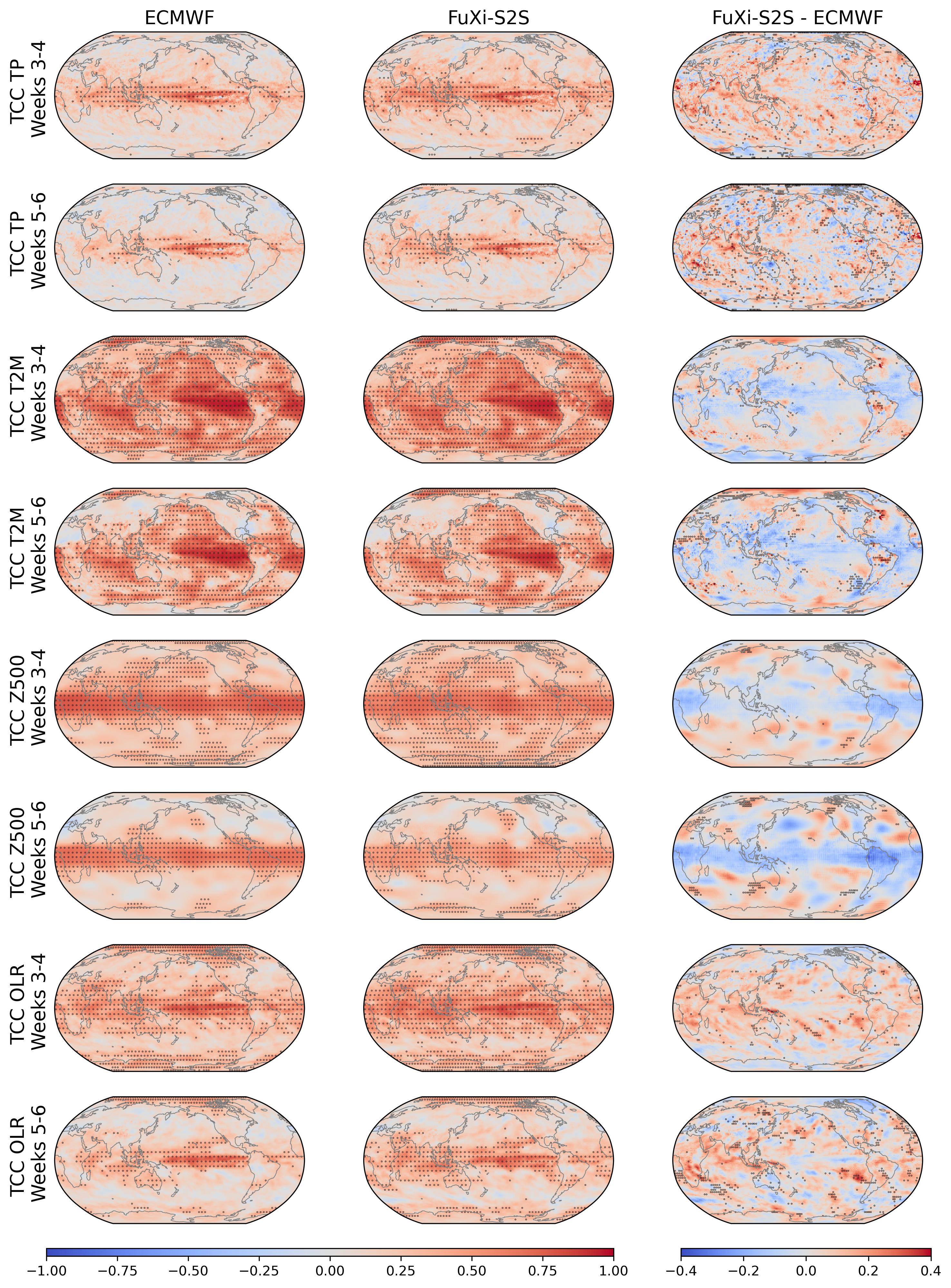}    
    \caption{Spatial map of average $\textrm{TCC}$ without latitude weighting of ECMWF S2S (first column) and FuXi-S2S (second column), and the differences in $\textrm{TCC}$ between FuXi-S2S and ECMWF S2S (third column) for $\textrm{TP}$ (first and second rows), $\textrm{T2M}$ (third and fourth rows), $\textrm{Z500}$ (fifth and sixth rows), and $\textrm{OLR}$ (seventh and eighth rows) at forecast lead times of weeks 3-4 (first, third, fifth, and seventh rows), weeks 5-6 (second, fourth, sixth, and eighth rows), using all testing data between 2017 and 2021. Stippling on the map denotes areas where the skill score is statistically significant at the 97.5\% confidence level. Specifically, in columns 1 and 2, stippling indicates regions where the skill scores of the ECMWF S2S and FuXi-S2S models significantly surpasses those of climatology. In column 3, stippling highlights areas where the FuXi-S2S model significantly outperforms the ECMWF S2S.}
    \label{TCC_spatial}    
\end{figure}
\FloatBarrier

\begin{figure}[h]
    \centering
    \includegraphics[width=\linewidth]{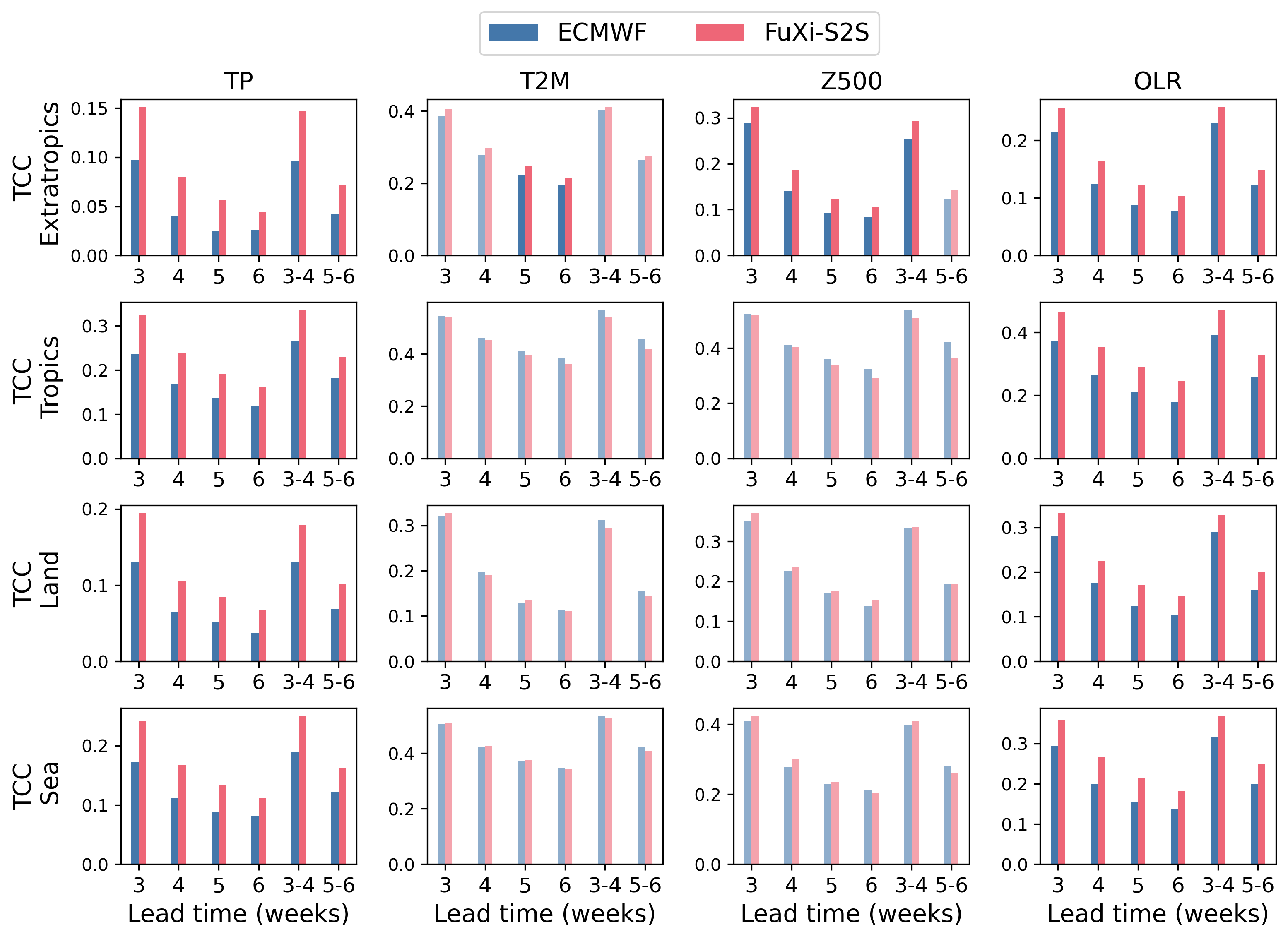}    
    \caption{Comparison of the latitude-weighted $\textrm{TCC}$ of the ensemble mean of ECMWF S2S (in blue) forecasts and FuXi-S2S forecasts (in red) for $\textrm{TP}$ (first column), $\textrm{T2M}$ (second column), $\textrm{Z500}$ (third column), and $\textrm{OLR}$ (fourth column) averaged over extra-tropics (90\textdegree S - 30\textdegree S and 30\textdegree N - 90\textdegree N, first row), tropics (30\textdegree S - 30\textdegree N, second row), land (third row), and sea (fourth row), using all testing data between 2017 and 2021. When the FuXi-S2S forecasts fail to show a statistically significant improvement over the ECMWF S2S reforecasts at the 97.5\% confidence level, a pale color scheme is used to denote these results.}
    \label{TCC_area_bar}    
\end{figure}
\FloatBarrier

\begin{figure}[h]
    \centering
    \includegraphics[width=\linewidth]{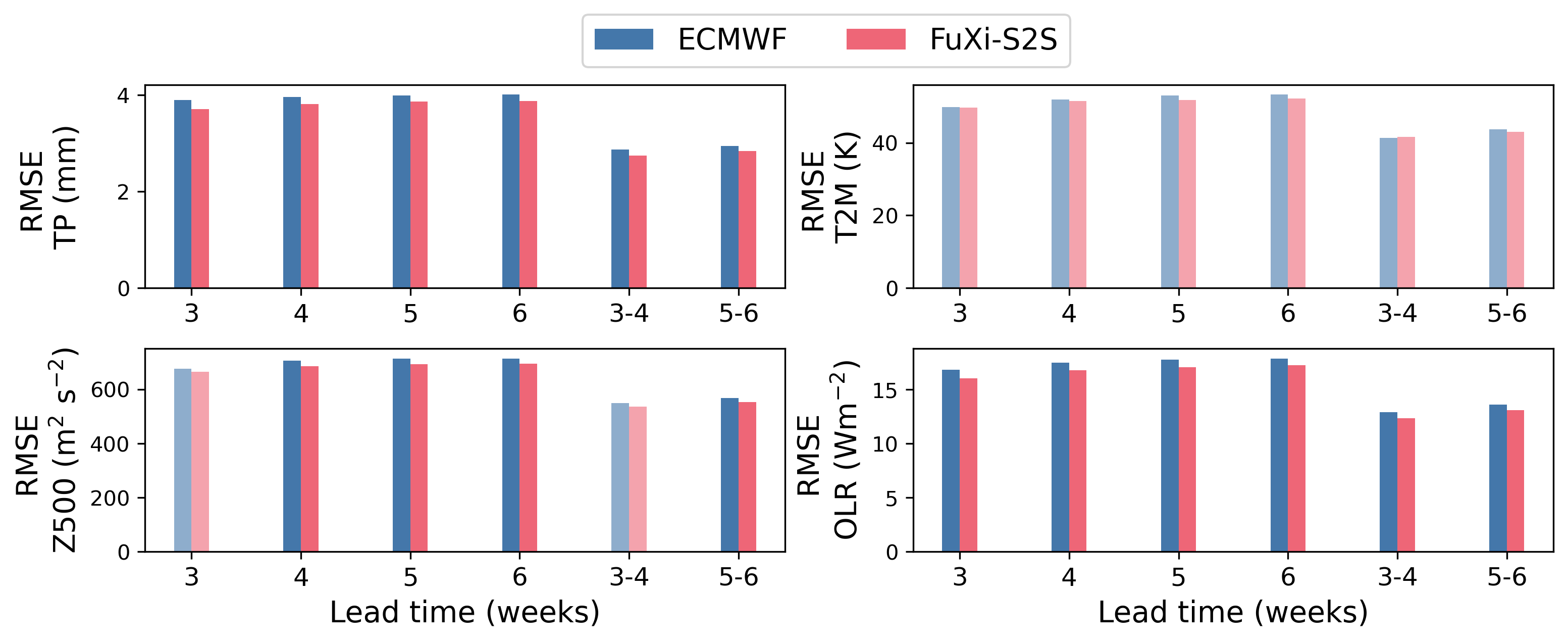}
    \caption{Comparison of the globally-averaged and latitude-weighted RMSE of the ensemble mean between ECMWF S2S reforecasts (in blue) and FuXi-S2S forecasts (in red) for $\textrm{TP}$, $\textrm{T2M}$, $\textrm{Z500}$, and $\textrm{OLR}$, using all testing data between 2017 and 2021. A bootstrapping approach, repeated 1000 times, is used for significance testing. When the FuXi-S2S forecasts fail to show a statistically significant improvement over the ECMWF S2S reforecasts at the 97.5\% confidence level, a pale color scheme is used to denote these results. It is important to note that $\textrm{TP}$ here refers to 24-hour accumulated precipitation.}
    \label{rmse}        
\end{figure}
\FloatBarrier

\begin{figure}
    \centering
    \includegraphics[width=\linewidth]{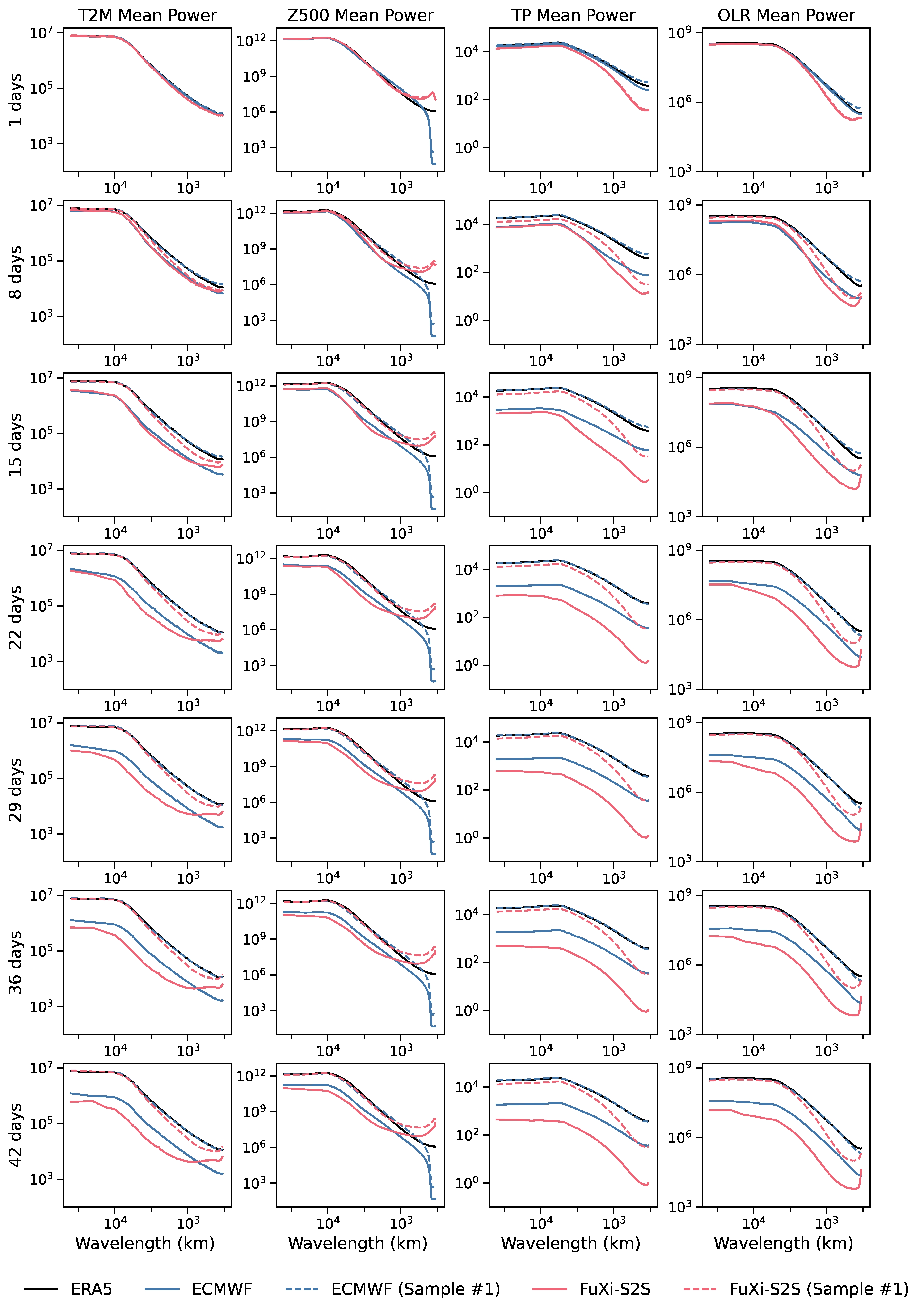}
    \caption{Energy spectra of $\textrm{T2M}$ (first row) and $\textrm{TP}$ (second row) for ERA5 (in black), ECMWF S2S (in blue) reforecasts and FuXi-S2S forecasts (in red) at forecast lead times of 15 days (first column), 22 days (second column), 29 days (third column), 36 days (fourth column), and 42 days (fifth column), using all testing data between 2017 and 2021.}
    \label{spectra}    
\end{figure}
\FloatBarrier

\begin{figure}[h]
    \centering
    \includegraphics[width=\linewidth]{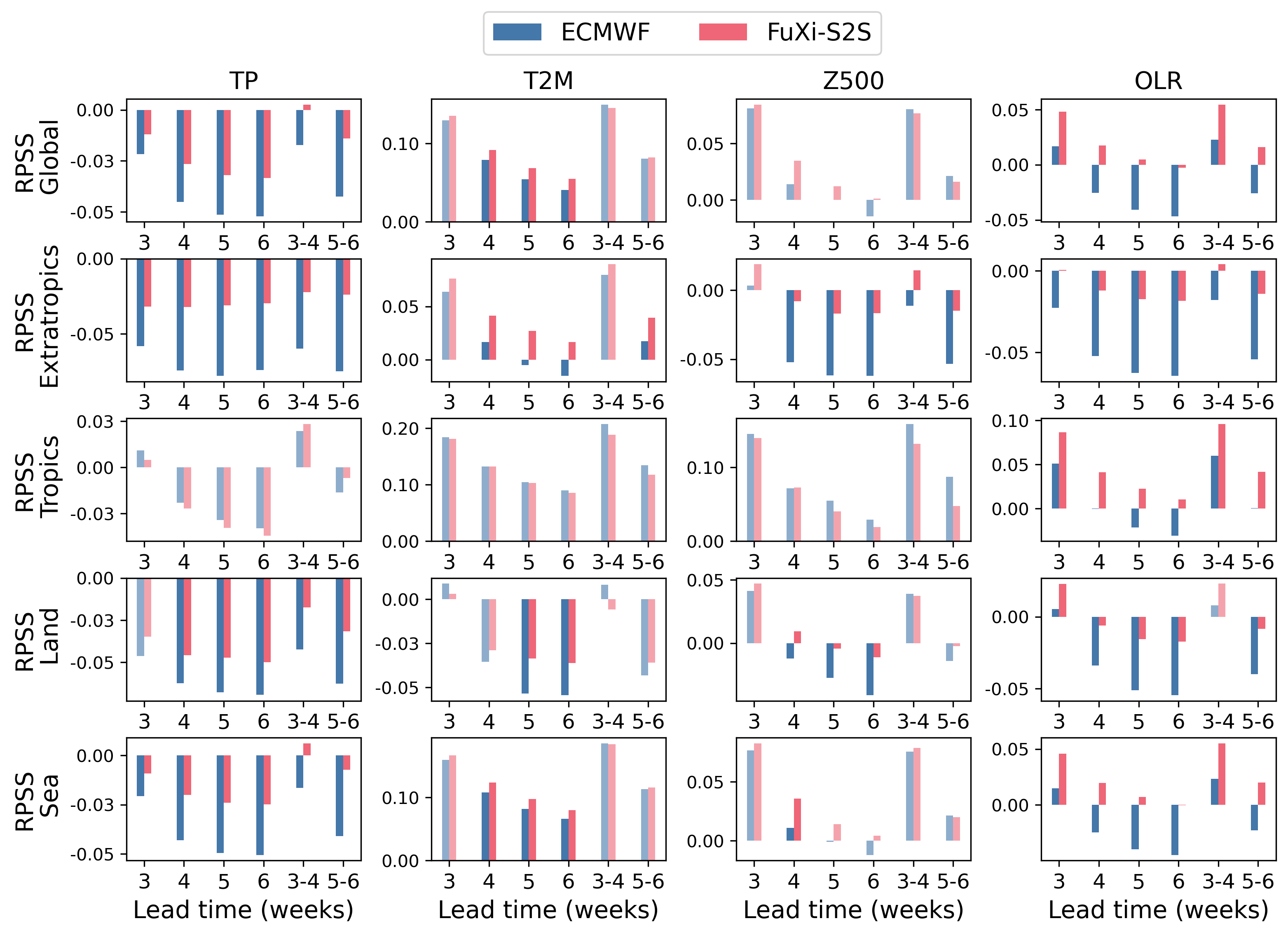}
    \caption{Comparison of the latitude-weighted $\textrm{RPSS}$ of ECMWF S2S (in blue) forecasts and FuXi-S2S forecasts (in red) for $\textrm{TP}$ (first column), $\textrm{T2M}$ (second column), $\textrm{Z500}$ (third column), and $\textrm{OLR}$ (fourth column) averaged over extra-tropics (90\textdegree S - 30\textdegree S and 30\textdegree N - 90\textdegree N, first row), tropics (30\textdegree S - 30\textdegree N, second row), land (third row), and sea (fourth row), using all testing data between 2017 and 2021. When the FuXi-S2S forecasts fail to show a statistically significant improvement over the ECMWF S2S reforecasts at the 97.5\% confidence level, a pale color scheme is used to denote these results.}
    \label{RPSS_area_bar}    
\end{figure}
\FloatBarrier

\begin{figure}[h]
    \centering
    \includegraphics[width=\linewidth]{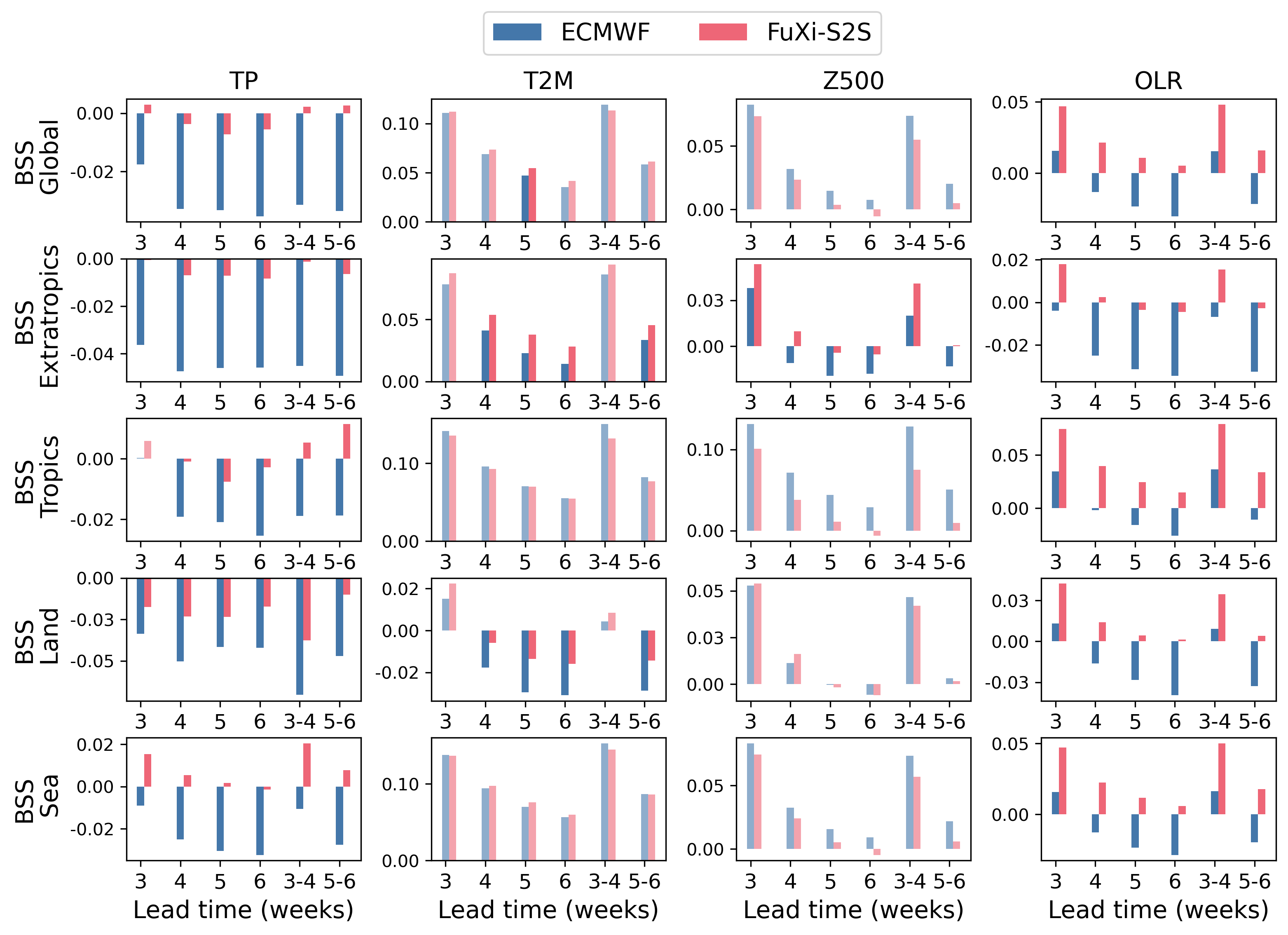}
    \caption{Comparison of the latitude-weighted $\textrm{BSS}$ of the ensemble mean of ECMWF S2S (in blue) forecasts and FuXi-S2S forecasts (in red) for $\textrm{TP}$ (first column), $\textrm{T2M}$ (second column), $\textrm{Z500}$ (third column), and $\textrm{OLR}$ (fourth column) averaged over extra-tropics (90\textdegree S - 30\textdegree S and 30\textdegree N - 90\textdegree N, first row), tropics (30\textdegree S - 30\textdegree N, second row), land (third row), and sea (fourth row), using all testing data between 2017 and 2021. When the FuXi-S2S forecasts fail to show a statistically significant improvement over the ECMWF S2S reforecasts at the 97.5\% confidence level, a pale color scheme is used to denote these results.}
    \label{BSS_area_bar}
\end{figure}
\FloatBarrier

\begin{figure}[h]
    \centering
    \includegraphics[width=\linewidth]{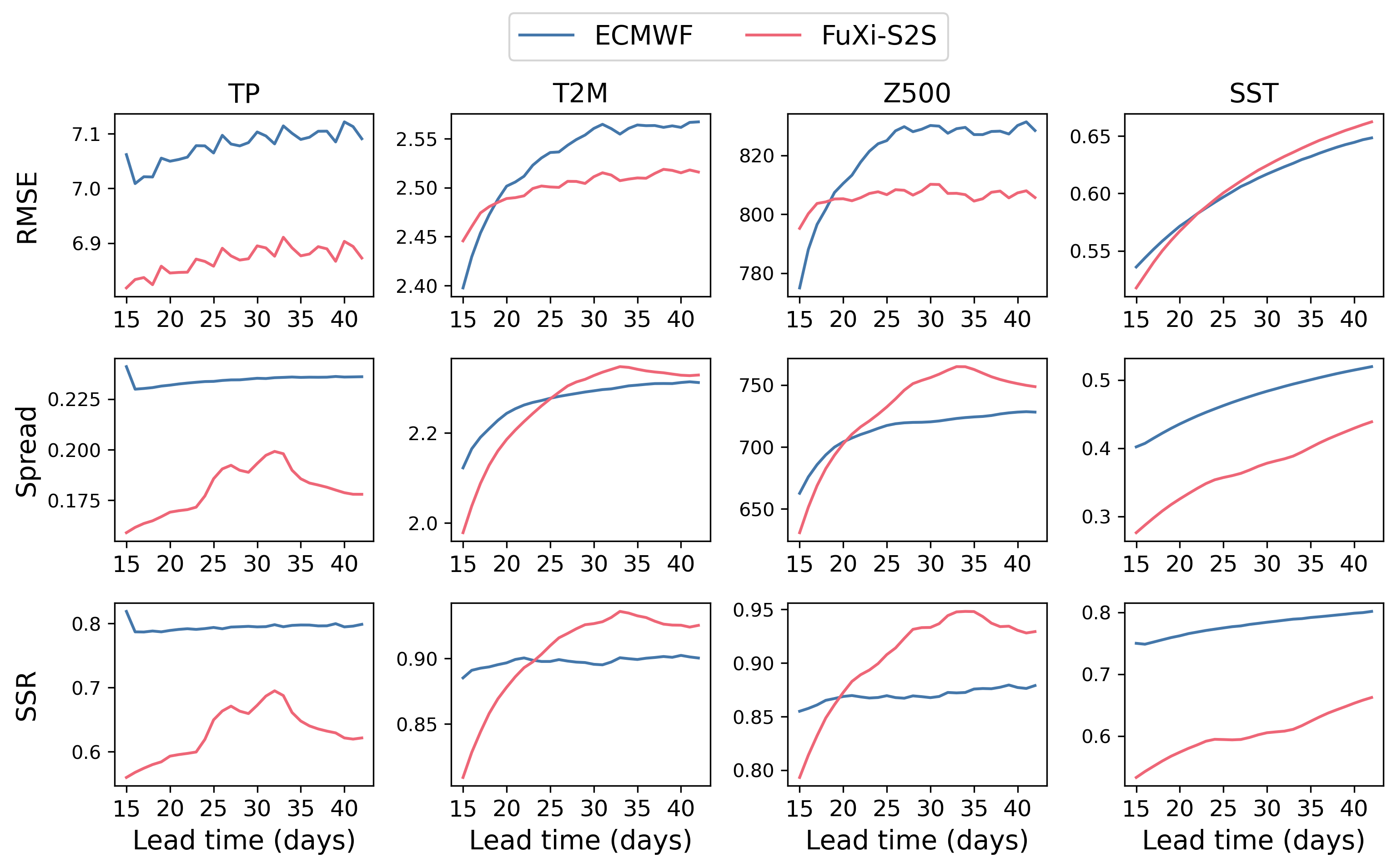}
    \caption{Comparison of the globally-averaged, latitude-weighted $\textrm{RMSE}$, ensemble spread, and $\textrm{SSR}$ of ECMWF S2S reforecasts (in blue), and FuXi-S2S forecasts (in red) for $\textrm{TP}$, $\textrm{T2M}$, and $\textrm{SST}$ as a function of forecast lead times. This analysis includes all testing data between 2017 and 2021, using the daily mean forecasts to calculate these metrics. When the FuXi-S2S forecasts fail to show a statistically significant improvement over the ECMWF S2S reforecasts at the 97.5\% confidence level, a pale color scheme is used to denote these results.}
    \label{spread_skill_plot}    
\end{figure}
\FloatBarrier

\begin{figure}[h]
    \centering
    \includegraphics[width=\linewidth]{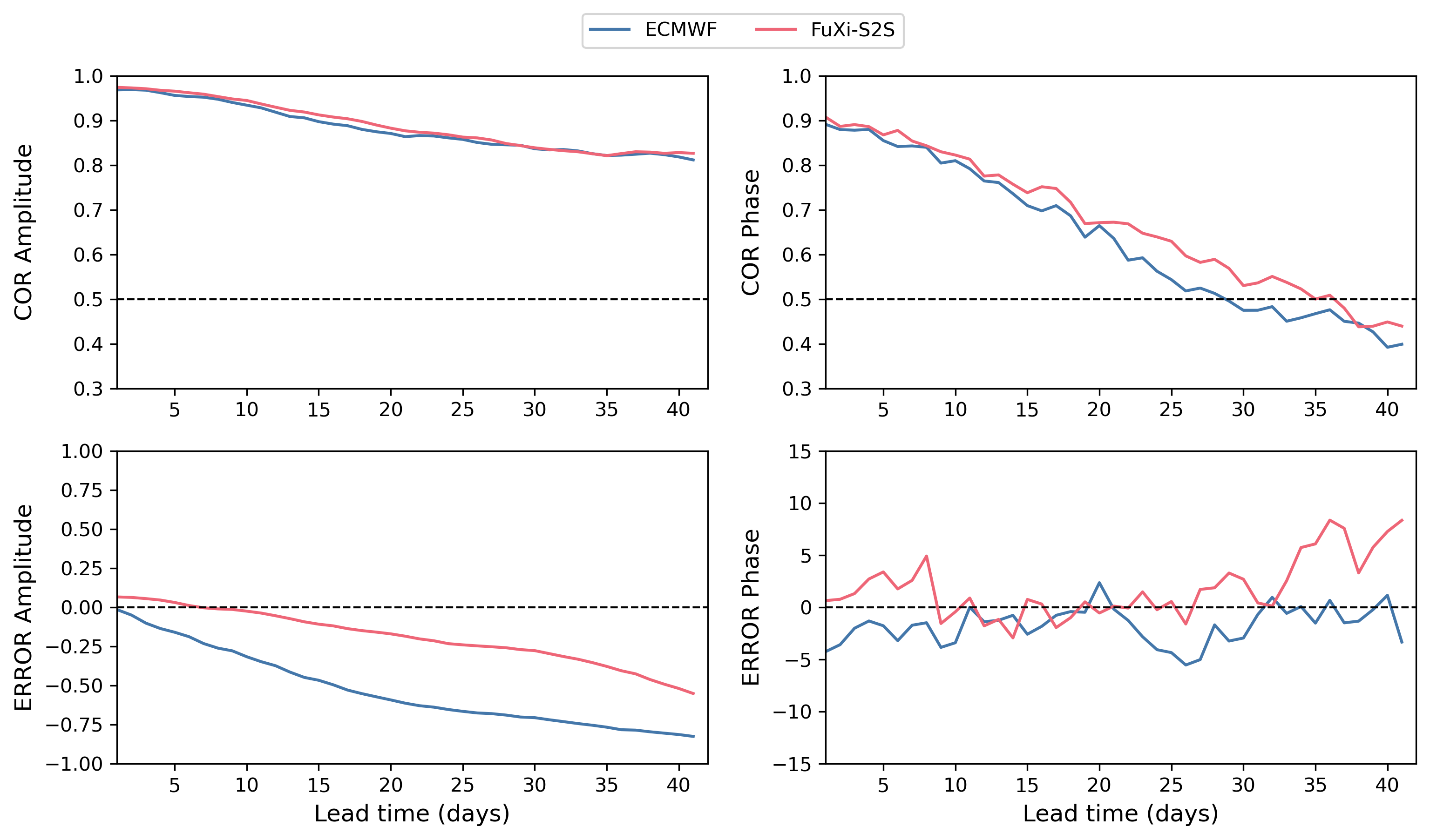}
    \caption{Comparison of correlation ($\textrm{COR}$) (first row) and error ($\textrm{ERROR}$) (second row) in the amplitude (first column) and phase (second column) of the MJO of the ensemble mean between ECMWF S2S reforecasts (in blue) and FuXi-S2S forecasts (in red) using all testing data from 2017 to 2021. Dashed black line signifies the prediction skill threshold of $\textrm{COR}$=0.5 and $\textrm{ERROR}$=0.}
    \label{mjo_amp_phase}    
\end{figure}
\FloatBarrier

\begin{figure}[h]
    \centering
    \includegraphics[width=\linewidth]{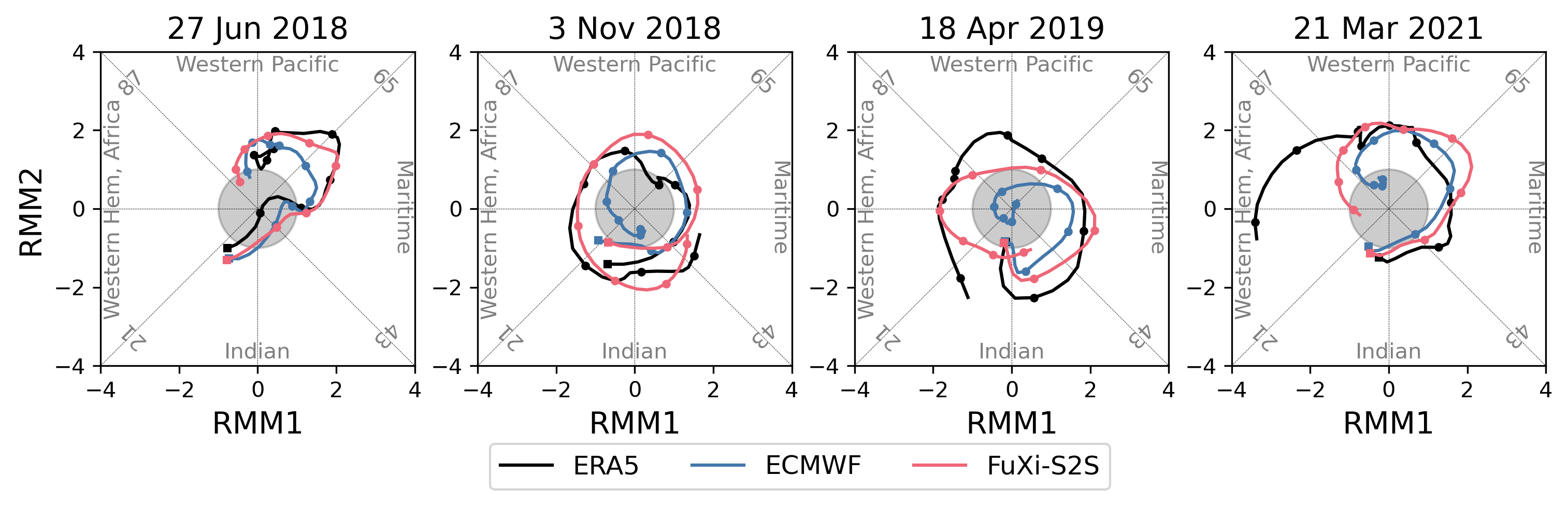}
    \caption{Comparison of the RMM composite phase–space diagram for the observed MJO derived from the combination of CBO and ERA5 reanalysis data (in black) and the ensemble mean of ECMWF S2S reforecasts (in blue), and FuXi-S2S forecasts (in red). RMM1 and RMM2 are the x axis and y axis, respectively. The numbers within each octant (from 1 to 8) are the defined MJO phase, and the words on each side of the diagram describe the approximate location of MJO associated convection along the equator. Squares represent forecasts on day 1 and closed circles represent every 5 days from the forecast initialization time (open squares). The panels are for different initialization date: 27 June 2018, 3 November 2018, 18 April 2019, and 21 March 2021.}
    \label{RMM_composite}    
\end{figure}
\FloatBarrier

\begin{figure}[h]
    \centering
    \includegraphics[width=\linewidth]{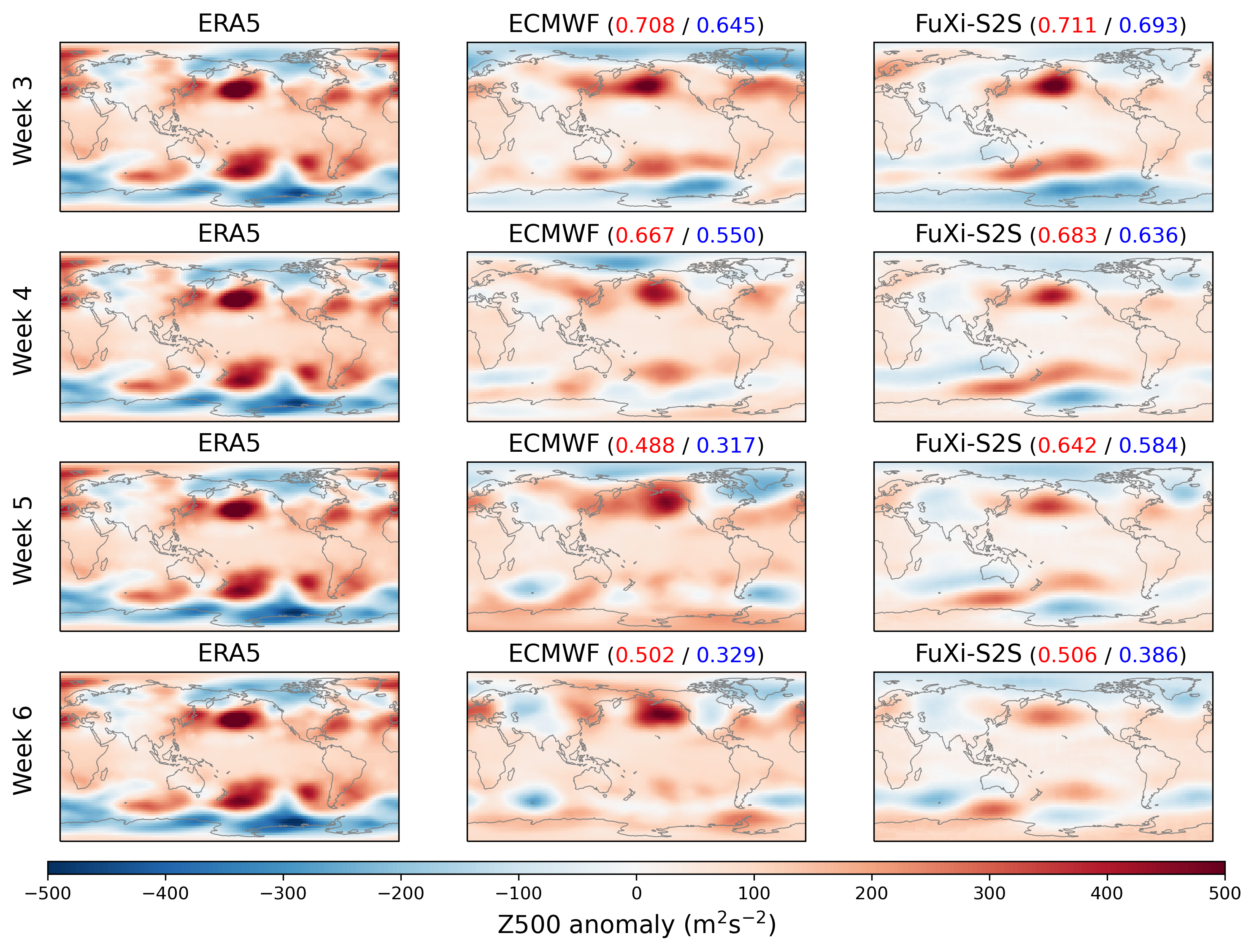}    
    \caption{Composite map of ${\textrm{Z500}}$ anomalies derived from ERA5 reanalysis data (first column), ECMWF S2S reforecasts (second column) and FuXi-S2S forecasts (third column). These maps cover forecast lead times of weeks 3, 4, 5, and 6, represented in the first, second, third, and fourth rows, respectively. All maps use testing data between 2017 and 2021, corresponding to initial forecast periods when the MJO is in phase 4 of its lifecycle. Red and blue numbers in columns 2 and 3 represent latitude-weighted pearson correlation coefficient (PCC) averaged globally and over extra-tropics (90\textdegree S - 30\textdegree S and 30\textdegree N - 90\textdegree N), respectively.}
    \label{z500}    
\end{figure}
\FloatBarrier

\begin{figure}[h]
    \centering
    \includegraphics[width=\linewidth]{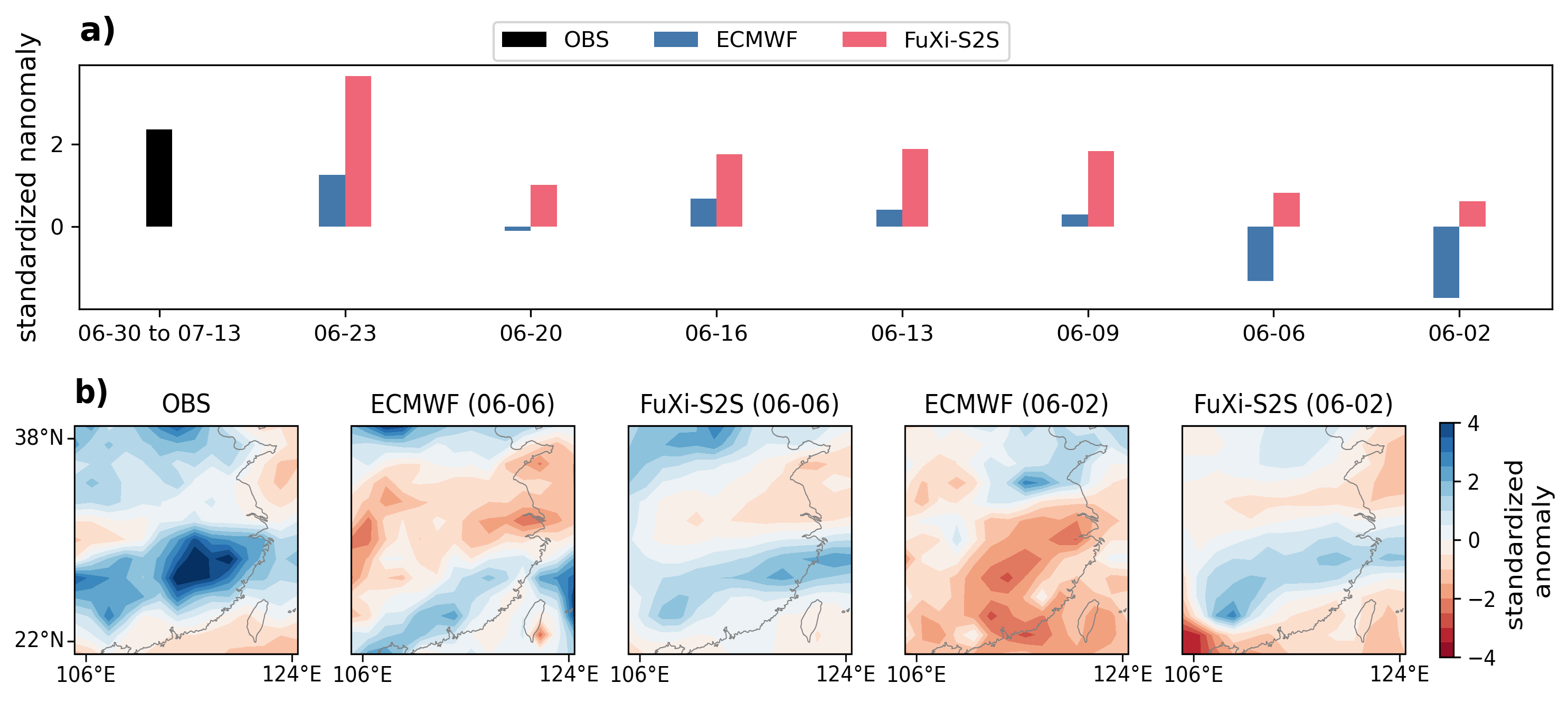}
    \caption{Comparison of the spatially and temporally averaged standardised $\textrm{TP}$ anomaly (a) for the 2 weeks from June 30th to July 13th, 2020 for GPCP observation (in black) and the predictions from ECMWF S2S reforecasts (in blue) and FuXi-S2S forecasts (in red), with initialization dates: June 23rd (06-23, MM-DD), June 20th (06-20), June 16th (06-16), June 13th (06-13), June 9th (06-09), June 6th (06-06), and June 2nd (06-02). Comparison of the temporally averaged standardised $\textrm{TP}$ anomaly maps (b) for GPCP observation (first column) and predictions from ECMWF S2S (second column) and FuXi-S2S (third column), with initialization dates on June 6th (06-06, first row), and June 2nd (06-02, second row).}
    \label{meiyu}    
\end{figure}
\FloatBarrier

\begin{figure}[h]
    \centering
    \includegraphics[width=\linewidth]{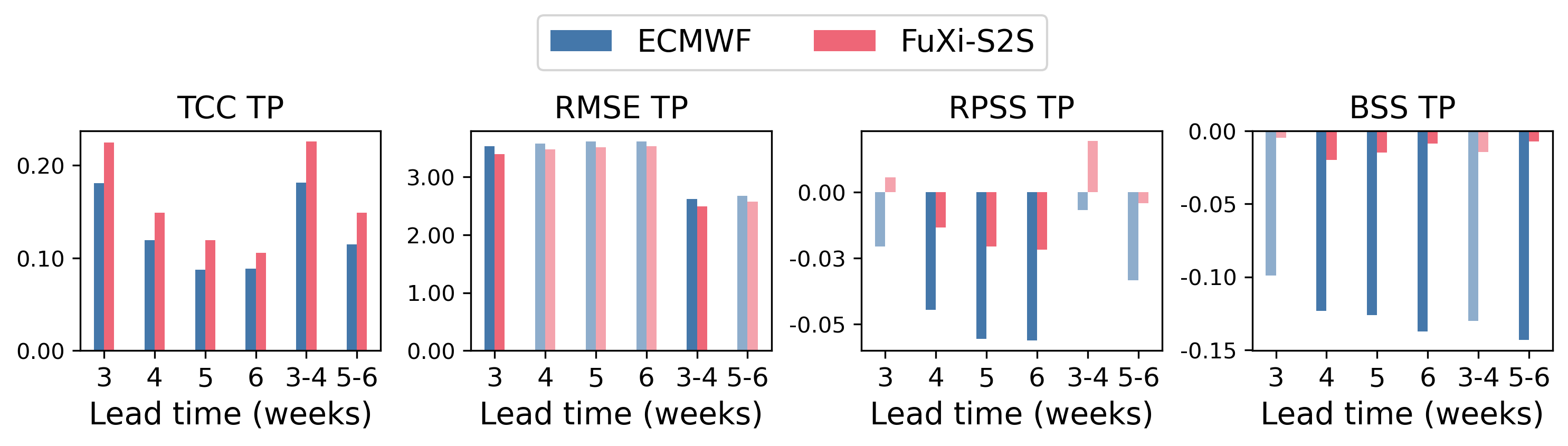}
    \caption{Comparison of the globally-averaged latitude-weighted TCC (first column), RMSE (second column), RPSS (third column), and BSS (fourth column) between ECMWF S2S real-time forecasts (in blue) and FuXi-S2S forecasts (in red) for $\textrm{TP}$, using testing data from 2022. When the FuXi-S2S forecasts do not demonstrate a statistically significant improvement over the ECMWF S2S reforecasts, a pale color scheme is used to denote these results. It is important to note that $\textrm{TP}$ here refers to 24-hour accumulated precipitation. When the FuXi-S2S forecasts fail to show a statistically significant improvement over the ECMWF S2S reforecasts at the 97.5\% confidence level, a pale color scheme is used to denote these results.}
    \label{tp_2022}    
\end{figure}
\FloatBarrier

\begin{figure}[h]
    \centering
    \includegraphics[width=\linewidth]{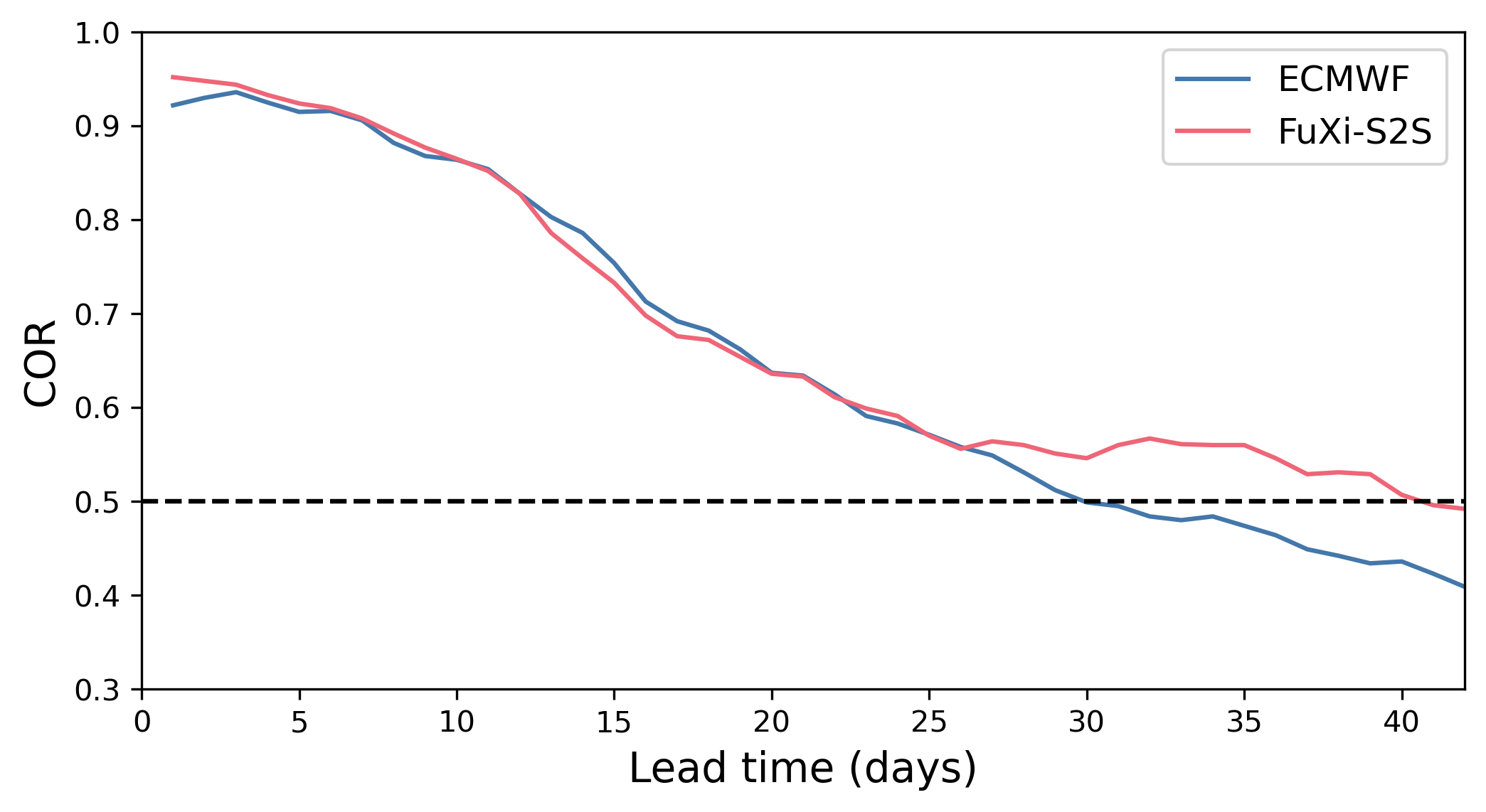}
    \caption{Comparison of the globally-averaged latitude-weighted RMM bivariate $\textrm{COR}$ (left column) of the ensemble mean of ECMWF S2S real-time forecasts (in blue) and FuXi-S2S forecasts (in red) using testing data from 2022, with dashed black lines indicating the prediction skill threshold of COR=0.5.}
    \label{MJO_2022}    
\end{figure}
\FloatBarrier

\begin{figure}[h]
    \centering
    \includegraphics[width=\linewidth]{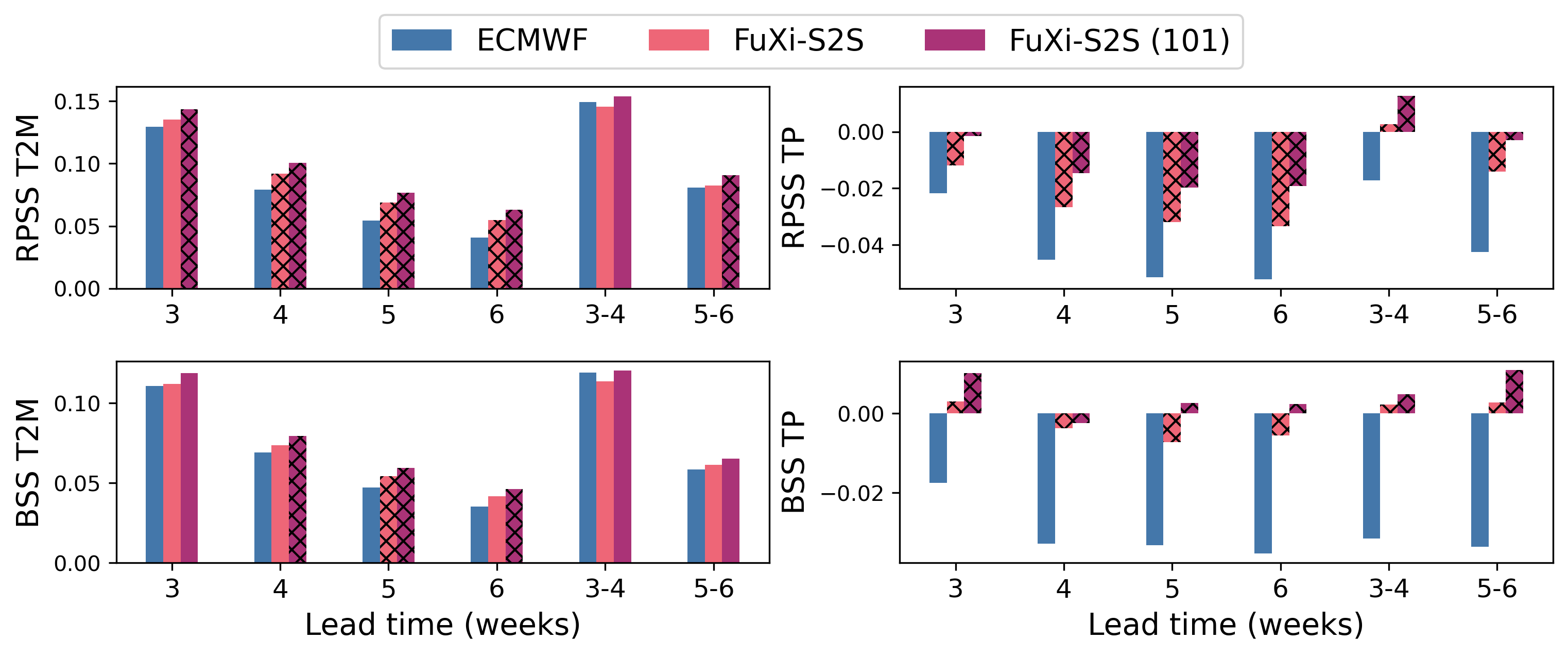}
    \caption{Comparison of the globally-averaged latitude-weighted RPSS (first row) and BSS (second row) between ECMWF S2S reforecasts (in blue), 51-member FuXi-S2S forecasts (in red), and 101-member FuXi-S2S forecasts (in purple) for $\textrm{T2M}$ and $\textrm{TP}$, using testing data from 2017 to 2021. When the 51-member FuXi-S2S forecasts or 101-member FuXi-S2S forecasts demonstrate a statistically significant improvement over the ECMWF S2S reforecasts at the 97.5\% confidence level, a cross-line on the bar plot is used to denote these results.}
    \label{larger_member}    
\end{figure}
\FloatBarrier

\begin{figure}[h]
    \centering
    \includegraphics[width=\linewidth]{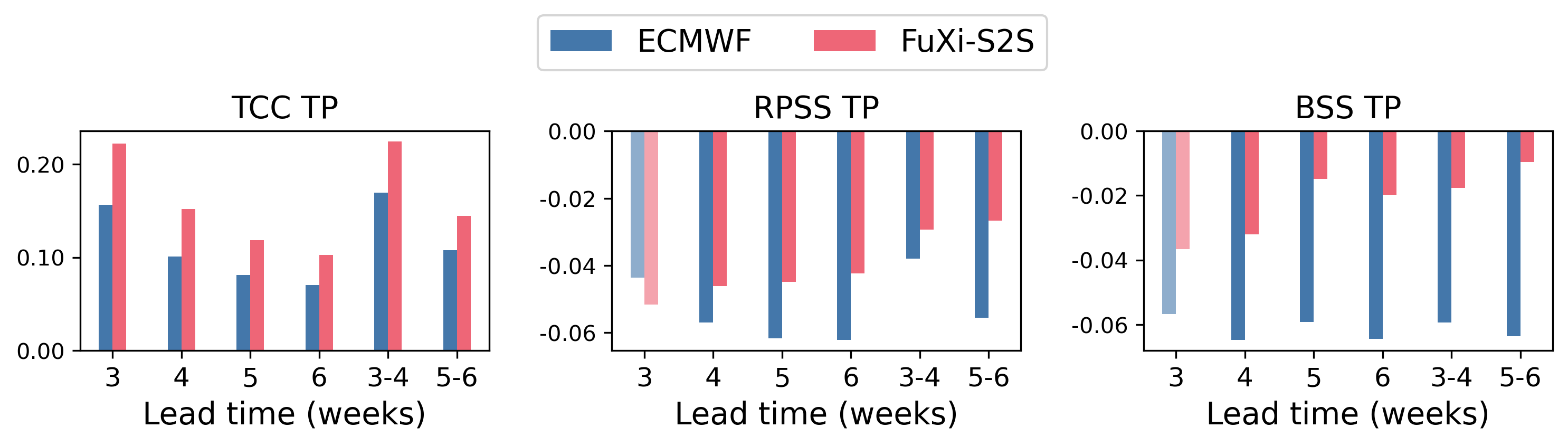}
    \caption{Comparison of the globally-averaged and latitude-weighted TCC, RPSS, and BSS between ECMWF S2S reforecasts (in blue) and FuXi-S2S forecasts (in red) for $\textrm{TP}$, using all testing data between 2017 and 2021. Notably, verification is conducted with the GPCP dataset, rather than ERA5 dataset. When the FuXi-S2S forecasts fail to show a statistically significant improvement over the ECMWF S2S reforecasts at the 97.5\% confidence level, a pale color scheme is used to denote these results.}
    \label{gpcp_tp}        
\end{figure}
\FloatBarrier

\section*{Supplementary Tables}

\begin{table}
\centering
\caption{\label{hyper} Optimizer hyperparameters}
\begin{tabular}{l|l}
\hline
Optimizer & AdamW \\
LR decay schedule & Cosine \\
Number of GPUs used & 8 \\
Batch size & 1 per GPU \\
Peak LR & 2.5e-4 \\
Weight decay & 0.1 \\
Total training steps  & 17,000 \\
\hline
\end{tabular}
\end{table}

\noindent

\bibliographystylesupp{sn-mathphys}
\bibliographysupp{refs}% common bib file
%\bibliography{refs}% common bib file

\end{CJK*}

\clearpage